\documentclass[preprint,10pt,3p]{elsarticle}

\journal{Computer Networks}

\usepackage{amsmath,amssymb}
\usepackage{graphicx}
\usepackage{booktabs}
\usepackage{multirow}
\usepackage{xurl}
\usepackage{xcolor}
\usepackage{tabularx}
\usepackage{longtable}
\usepackage{caption}
\usepackage{array}
\usepackage{algorithm}
\usepackage{algpseudocode}
\usepackage[hidelinks]{hyperref}

\newcommand{\plainid}[1]{%
  \ifmmode
    \mathrm{#1}%
  \else
    \textrm{#1}%
  \fi
}

\begin{document}

\begin{frontmatter}

\title{Authority-Bound Governance of Heterogeneous AI Security Decisions in Telecom and IoT Networks}

\author[inst1]{Saviz Changizi}
\author[inst2]{Nasibeh Mohammadzadeh}
\author[inst1]{Mohammad Shojafar}
\author[inst1]{Rahim Tafazolli}

\affiliation[inst1]{
organization={6G Innovation Centre, Institute for Communication Systems, University of Surrey},
city={Guildford},
country={United Kingdom}
}

\affiliation[inst2]{
organization={Department of Network Engineering, Universitat Polit\`ecnica de Catalunya (UPC)},
city={Barcelona},
country={Spain}
}

\begin{abstract}
Artificial intelligence (AI)-enabled security decision systems in telecom and IoT networks can draw on heterogeneous models whose outputs may trigger operational actions. Recording such a decision on a blockchain, however, does not establish that it is authorised, applicable, policy-consistent, or still valid when execution is attempted. This paper presents governance-2, an authority-bound and fail-closed governance architecture that separates upstream scientific decision formation from downstream operational enforcement of network security decisions. Each governed case is bound to registered dataset, model, policy, deployment, and optional refiner authorities, while smart-contract checks enforce role separation, authority compatibility, score-to-state and state-to-action consistency, lifecycle validity, replay protection, pause control, authority revocation, and exact-action execution. Evaluation uses two independent branches: a controlled spectrum-access replay with four frozen heterogeneous decision configurations and a measured radio-frequency branch based on WiFiSpectralJam. Across eight frozen measured-data decision streams, the governance layer processes 153,744 stream--case instances derived from 19,218 measured captures while preserving interference-specific semantics and unresolved review states. The full contract rejects all tested invalid operations; randomised stateful invariant testing completes 2,000 generated transaction sequences with zero invariant violations; and single-capability ablation shows that removing an enforcement family exposes its assigned invalid operations while unrelated protections remain active and legitimate control lifecycles remain executable. The results indicate that heterogeneous AI security decisions can share a common governance plane across telecom and IoT network settings without redefining their scientific semantics, while explicit authority and lifecycle enforcement constrains which decisions are permitted to reach operational execution.
\end{abstract}


\end{frontmatter}

\section{Introduction}
\label{sec:introduction}

AI-supported security decision systems are used in telecommunications networks \cite{edozie2025telecomai} and Industrial Internet of Things (IIoT) environments \cite{raeiszadeh2024iiot}. Telecommunications fraud is one prominent operational challenge. The Communications Fraud Control Association (CFCA) estimated global telecommunications fraud losses at USD~38.95 billion in 2023, equivalent to 2.5\% of global telecommunications revenues \cite{cfca2023}. Reporting based on the 2025 CFCA Global Fraud Loss Survey places the corresponding loss at approximately USD~41.82 billion \cite{tns2026}. The same report states that 75\% of surveyed operators use machine learning (ML) or artificial intelligence (AI) for fraud detection \cite{tns2026}.

Security decisions in telecom and IoT networks can be produced from markedly different forms of evidence and learning architectures. Spatial--temporal modelling combines sequential and structural calling behaviour for telecom-fraud detection \cite{chu2023telecomfraud}, whereas latent-synergy graph learning captures coordinated fraud relationships across users \cite{wu2024latent}. Large language models (LLMs) have been applied to structured semantic classification of telecom-fraud text \cite{ding2026mrml}. Federated learning (FL) provides a distributed learning setting for 6G security, but also introduces security risks associated with collaborative model training \cite{dealwis2026fl6g}. These examples illustrate a wider operational problem: network-security decisions may originate from different models, evidence types, and decision processes, yet may ultimately enter the same operational control workflow.

Blockchain addresses a different part of this problem. In communication networks, it has been used for functions including secure records, decentralised identity, settlement, resource trading, network slicing, and edge coordination \cite{mamun2026blockchaincomms}. Blockchain-enabled IoT and IIoT intrusion-detection systems have been organised around rule-based, ML, deep-learning, and FL approaches according to the objects placed or governed on-chain \cite{chen2026blockchainids}. In 6G-oriented architectures, blockchain has also been considered alongside fog and edge computing and FL as part of broader security and privacy designs \cite{valdez2026sixg}. These uses show the value of blockchain for distributed trust and control, but they do not by themselves determine whether a specific AI-generated security decision is authorised to become an operational network action.

The transition from an already-formed security decision to operational execution is a distinct control boundary. A hybrid telecom-fraud framework has connected ML- and LLM-derived decisions to blockchain-supported enforcement \cite{changizi2025hybrid}. Trust-aware Internet of Medical Things (IoMT) access control has similarly combined off-chain ML with blockchain-backed access decisions and audit records \cite{alashwal2026iomt}. At execution time, however, recording or forwarding a decision is not sufficient. The decision may still need to be checked against the dataset and model context from which it originated, the policy authorised for that model, the action permitted by that policy, its current lifecycle state, and the continued validity of the authorities on which it depends.

This paper addresses that boundary through governance-2, an authority-bound and fail-closed governance architecture for heterogeneous AI security decisions in telecom and IoT networks. Scientific decision formation remains upstream, while the governance layer operates before operational execution. It consumes frozen decision evidence and binds each governed case to registered dataset, model, policy, deployment, and optional refiner authorities. The smart contract checks policy-defined score--state--action consistency, role authorisation, lifecycle validity, replay protection, pause status, authority revocation, refinement applicability, and the exact action authorised for execution. The purpose is not to reproduce or judge the upstream learning process, but to determine whether an already-formed decision is currently admissible for execution under the registered governance conditions.

The evaluation uses two independent security-decision branches. The controlled branch replays spectrum-access decisions produced by four frozen upstream configurations spanning centralised ML, federated meta-learning, and LLM-family decision sources. The measured-data branch uses WiFiSpectralJam, a commodity Wi-Fi spectral-scan dataset containing 96,090 files collected under benign and controlled radio-frequency (RF) jamming conditions \cite{herzalla2026wifispectraljam}. The RF branch retains interference-specific decision semantics and does not reinterpret measured jamming observations as telecom-fraud labels. Both branches use the same downstream governance mechanism while preserving their respective scientific meanings and operational actions.

The main contributions are:
\begin{itemize}
    \item an authority-bound governance architecture for telecom and IoT network-security decisions that binds heterogeneous AI outputs to explicit dataset, model, policy, deployment, and optional refiner authorities before operational execution;
    \item fail-closed enforcement of role authorisation, authority compatibility, score--state--action consistency, final-action integrity, lifecycle and replay safety, refiner applicability, administrative pause, and authority revocation;
    \item evaluation across two distinct security-decision settings using controlled spectrum-access decisions and independently measured RF-interference data, showing how a common governance plane can preserve domain-specific decision semantics and unsupported-review states; and
    \item adversarial, stateful-invariant, and single-capability-ablation evaluation of invalid governance transitions and multi-step lifecycle behaviour, complemented by legitimate lifecycle controls and measurements of prototype execution cost.
\end{itemize}

The remainder of the paper is organised as follows. Section~\ref{sec:related-work} reviews AI-based security decision formation and blockchain-supported security mechanisms. Section~\ref{sec:system-threat} defines the system and threat models. Section~\ref{sec:governance} presents the authority-bound governance architecture, and Section~\ref{sec:security-properties} describes its enforcement properties. Section~\ref{sec:experiments} details the experimental methodology, Section~\ref{sec:results} reports the results, and Section~\ref{sec:discussion} discusses the findings, limitations, and future directions. Section~\ref{sec:conclusion} concludes the paper.

\section{Related Work}
\label{sec:related-work}

\subsection{AI-Based Security Decision Formation}
\label{subsec:related-ai}

Sequential calling behaviour has been combined with graph-based structural patterns to derive user-level telecom-fraud scores \cite{chu2023telecomfraud}. Temporal bipartite calling sequences have also been modelled through a Hawkes-enhanced long short-term memory architecture that captures repetition and correlation in communication behaviour \cite{jiang2023hesm}. Coordinated fraud has been represented through latent-synergy graph learning combined with sequential behaviour \cite{wu2024latent}. Graph imbalance has been addressed through cost-sensitive graph attention so that minority fraudulent nodes receive greater emphasis during learning \cite{hu2024gatcobo}.

Wangiri detection has been studied under highly imbalanced and initially unlabeled call-detail-record data using domain-driven heuristic labelling followed by supervised model comparison \cite{balouchi2026wangiri}. LLM-based telecom-fraud classification has been structured through multi-role, multi-layer prompting \cite{ding2026mrml}. Intermediate-layer neuron selection has also been used to identify informative LLM representations rather than relying only on final-layer outputs \cite{jiang2025lens}. Broader telecom-anomaly research documents the movement from rule-based methods toward ML, deep learning, and adaptive AI approaches \cite{edozie2025telecomai}.

AI-based anomaly detection has been evaluated for cloudified mobile-core architectures \cite{michelinakis2023cloudcore}. Real-time IIoT anomaly detection has combined multi-source prediction with concept-drift adaptation \cite{raeiszadeh2024iiot}. FL has been studied both as a security mechanism and as an attack surface in 6G environments \cite{dealwis2026fl6g}. Poisoning defence in zero-touch Beyond-5G (B5G) management has included trusted-participant selection and identification of malicious model updates \cite{bensaad2023poisoning}.

Dynamic blockchain-fraud detection has been addressed through automated pattern discovery, hierarchical representation learning, heterogeneous graph transformers, and continual learning \cite{gao2026ahgt}. Heterogeneous source models and collaborative knowledge transfer have been used to improve fraud detection on resource- or data-limited transaction platforms \cite{tang2025gmdk}. A modular telecom-fraud framework has combined conventional ML, LLM reasoning, and blockchain smart contracts within the same fraud-control architecture \cite{changizi2025hybrid}.

\subsection{Blockchain-Supported Network Security, Trust, and Control}
\label{subsec:related-blockchain}

Blockchain applications in communication networks include decentralised identity, secure records, settlement, resource trading, and 5G/6G networking functions \cite{mamun2026blockchaincomms}. Blockchain-based dynamic spectrum access and network slicing have been studied as radio-resource-management and traceability problems \cite{muntaha2023dsa}. Spectrum sensing has also been integrated directly into blockchain consensus for detecting spectrum-access violations \cite{fernando2023dpos}. In 5G network slicing, blockchain has been used to support transparent resource exchange under front-running, selfish-validation, confidentiality, and trust constraints \cite{diallo2025trade5g}. Blockchain has also been integrated into federated encrypted-traffic classification, where validators assess local models and a quality-driven aggregation mechanism filters malicious or low-quality updates before the global model is uploaded to the blockchain \cite{liu2025qdvta}.

Decentralised reputation in B5G/6G environments has been implemented through smart-contract-based trust mechanisms and evaluated under trust-related attacks \cite{nunez2026nxgent}. For 6G trust management, FBBTrust combines fuzzy inference and BiLSTM-based temporal modelling with lightweight blockchain persistence for tamper-resistant trust updates and decentralised verification \cite{saeedi2026fbbtrust}. Cross-domain data sharing has been supported through sharding blockchain, policy registration, authentication, and access-control procedures under partial-trust and zero-trust assumptions \cite{liu2024crossdomain}. Social Internet of Things (SIoT) trust assessment has been derived from auditable on-chain interaction and latency evidence rather than application-level semantic correctness \cite{luo2026bctrust}. Blockchain has also been used as a decentralised trust base for protecting trusted execution environment (TEE) state continuity against rollback and forking attacks \cite{peng2024narrator}.

In IoMT access control, computationally intensive ML and explainability functions have been kept off-chain while access decisions and audit records are handled through a hybrid blockchain architecture \cite{alashwal2026iomt}. In SDN intrusion detection, BLOC-SHIELD combines trust-aware federated aggregation with a lightweight permissioned blockchain that records aggregation metadata as an auxiliary audit layer without directly participating in model training or inference \cite{ram2026blocshield}. Blockchain-integrated deep-learning intrusion-detection systems for industrial control environments have been reviewed in terms of real-time data availability, reproducibility, system effectiveness, computational complexity, and deployment cost \cite{devipriya2025ics}. Blockchain-enabled IoT/IIoT intrusion-detection research has been organised into trusted-rule, ML, deep learning, and FL paradigms according to the principal object placed or governed on-chain \cite{chen2026blockchainids}. A 6G-oriented taxonomy places blockchain alongside fog/edge computing and FL across security, privacy, data-management, policy, governance, and deployment dimensions \cite{valdez2026sixg}.

\subsection{Positioning of This Work}
\label{subsec:related-positioning}

Representative blockchain-supported systems address different control objects across network and security workflows. Spectrum sensing can be incorporated into the blockchain consensus process itself \cite{fernando2023dpos}. Blockchain can govern resource exchange within a 5G slicing marketplace \cite{diallo2025trade5g}. Smart contracts can govern decentralised reputation updates in B5G/6G environments \cite{nunez2026nxgent}. Cross-domain policy registration and access control can be enforced through sharding-blockchain protocols \cite{liu2024crossdomain}. Blockchain can provide a trust base for TEE state continuity \cite{peng2024narrator}. On-chain interaction telemetry can be used to produce auditable SIoT trust estimates \cite{luo2026bctrust}.

Other architectures place AI closer to the downstream control boundary. Off-chain AI has been used to contribute to trust-aware IoMT access decisions that are subsequently recorded through a hybrid ledger \cite{alashwal2026iomt}. Heterogeneous ML- and LLM-derived telecom-fraud decisions have been connected to smart-contract-supported enforcement and logging \cite{changizi2025hybrid}. A recent adaptive IoT IDS combines lightweight edge ML and LLM-based unknown-traffic analysis with a post-quantum-secure blockchain database for protecting shared IDS data and managing authority-node trust \cite{huang2026lqbids}. Blockchain-enabled intrusion detection system (IDS) architectures frequently anchor rules, model artefacts, alerts, or collaborative-learning evidence on-chain \cite{chen2026blockchainids}. Blockchain-assisted deep-learning IDS research remains primarily centred on intrusion detection and the practical costs of integrating ledger and detection components \cite{devipriya2025ics}. Policy and governance appear as broader design dimensions in 6G architectures that combine fog/edge computing, blockchain, and FL \cite{valdez2026sixg}.

Governance-2 starts after the scientific decision evidence has been formed and focuses on the point at which an AI-derived security decision can progress toward operational execution in a telecom or IoT network. Its novelty does not rest on role control, hash-based provenance, or lifecycle state machines as isolated mechanisms. Rather, the governed object is the already-formed AI decision, whose progression toward execution is conditioned on compatible dataset, model, policy, and optional refiner authorities together with the deployment context. These authority conditions are enforced alongside policy-defined decision consistency, applicability, role permissions, lifecycle state, pause status, authority revocation, and exact-action execution at the corresponding governance transition. The contribution is an authority-bound execution boundary for network-security decisions, determining whether an already-formed decision is currently authorised to progress toward an operational action rather than merely recording that decision or governing the upstream learning process.

Table~\ref{tab:related-work-positioning} compares representative systems using the downstream network-security decision-execution boundary evaluated in this work.

\begin{table}[!htbp]
\caption{Positioning of representative related systems against the downstream network-security decision-execution boundary considered in this work.}
\label{tab:related-work-positioning}
\centering
\scriptsize
\renewcommand{\arraystretch}{1.12}
\setlength{\tabcolsep}{3pt}

\begin{tabularx}{\textwidth}{
    c
    >{\raggedright\arraybackslash}X
    >{\raggedright\arraybackslash}X
    >{\centering\arraybackslash}p{1.45cm}
    >{\centering\arraybackslash}p{1.70cm}
    >{\centering\arraybackslash}p{1.70cm}
    >{\centering\arraybackslash}p{1.55cm}
}
\toprule
\textbf{Ref.} &
\textbf{Primary protected object} &
\textbf{Blockchain role} &
\textbf{AI decision role} &
\textbf{AI authority binding} &
\textbf{AI decision lifecycle} &
\textbf{Governance ablation} \\
\midrule

\cite{gao2026ahgt} &
Fraud prediction &
Target security domain &
Produces &
NR &
NR &
NR \\

\cite{liu2025qdvta} &
Encrypted traffic classification &
FL model exchange and recording &
Produces &
NR &
NR &
Yes (component-level only) \\

\cite{fernando2023dpos} &
Spectrum-access compliance &
Consensus and sensing &
None &
NR &
NR &
NR \\

\cite{diallo2025trade5g} &
5G resource exchange &
Marketplace enforcement &
None &
NR &
NR &
NR \\

\cite{liu2024crossdomain} &
Cross-domain data sharing &
Policy and access control &
None &
NR &
NR &
NR \\

\cite{peng2024narrator} &
TEE state continuity &
Decentralised trust base &
None &
NR &
NR &
NR \\

\cite{luo2026bctrust} &
SIoT trust assessment &
Auditable telemetry &
None &
NR &
NR &
NR \\

\cite{saeedi2026fbbtrust} &
6G node trust &
Trust-update persistence and verification &
Produces &
NR &
NR &
Yes (Component-level only) \\

\cite{alashwal2026iomt} &
IoMT access control &
Decision and audit recording &
Produces &
NR &
NR &
NR \\

\cite{ram2026blocshield} &
SDN federated intrusion detection &
Aggregation-metadata audit &
Produces &
NR &
NR &
NR \\

\cite{changizi2025hybrid} &
Telecom fraud control &
Decision enforcement and logging &
Produces &
NR &
NR &
NR \\

\cite{huang2026lqbids} &
IoT intrusion detection &
Data integrity and authority-node trust &
Produces &
NR &
NR &
Yes (component-level only) \\

\textbf{Our work} &
\textbf{AI decision execution} &
\textbf{Verification and enforcement} &
\textbf{Consumes} &
\textbf{Yes} &
\textbf{Yes} &
\textbf{Yes} \\

\bottomrule
\end{tabularx}

\vspace{1mm}
\begin{minipage}{\textwidth}
\scriptsize
\textit{Note:} NR indicates that the corresponding AI-decision governance mechanism is not reported as an explicit feature of the reviewed design; it does not imply the absence of other security or governance mechanisms in that work. AI authority binding refers specifically to explicit dataset--model--policy binding for an AI decision. ``Component-level only'' denotes ablation of predictive, trust, or aggregation components rather than ablation of downstream decision-governance enforcement properties.
\end{minipage}

\end{table}

\section{System Model and Threat Model}
\label{sec:system-threat}

\subsection{System Model}
\label{subsec:system-model}
The proposed system separates upstream scientific decision formation from downstream governance enforcement. Upstream AI components generate security-related decision evidence according to their own frozen data, model, and operating-policy authorities. The governance layer does not retrain models, select thresholds, redefine labels, or create new decision semantics. Instead, it consumes already-formed decision evidence and determines whether that evidence is authorised and executable under the registered governance rules.

The system supports heterogeneous security-decision domains through a shared governance plane. In the synthetic branch, governed cases represent spectrum-access security decisions, whereas the real-data branch uses measured Wi-Fi spectral captures from WiFiSpectralJam and represents RF-interference events. These domains retain their native decision semantics while sharing the same governance primitives for provenance, authority verification, lifecycle control, and auditable execution.

Each governed case is associated with a dataset authority, model authority, policy authority, deployment context, decision state, and requested action. Where automated refinement is applicable, a separate refiner authority may also be registered. The governance contract verifies that the submitted evidence is consistent with the registered authorities and with the case-specific policy before an action can reach authorised execution.

Operational responsibilities are separated across governance roles. Validators submit initial decisions, reviewers resolve cases requiring review, executors perform authorised actions, and authority administration is restricted to designated administrative roles. Governed cases progress through an explicit lifecycle comprising submission, optional review, decision readiness, execution, and finalisation. Invalid authority bindings, inconsistent decision semantics, unauthorised actions, or invalid lifecycle transitions are rejected rather than silently recorded.

\subsection{Trust Assumptions and Adversary Model}
\label{subsec:threat-model}
The governance mechanism controls the transition from an already-formed AI security decision to an authorised operational action. It does not treat the blockchain as an oracle for predictive truth. In particular, the contract does not establish that a registered model actually generated a submitted score, nor does it independently verify the correctness of upstream inference. Instead, it verifies whether submitted decision evidence is bound to registered authorities and is consistent with the frozen policy, applicability, role, and lifecycle rules governing the corresponding case.

\textbf{Trust assumptions.}
We assume that the governance contract is deployed from the intended implementation and that the underlying blockchain execution environment correctly enforces the deployed smart-contract logic. The cryptographic hash functions used to bind external evidence and registered authorities are assumed to provide their standard integrity properties. Authority registration and revocation are privileged administrative operations, so the authority administrator is trusted to register the intended dataset, model, policy, and, where applicable, refiner authorities. Compromise of this administrative trust root is outside the present threat model. Upstream scientific artefacts, including model outputs, thresholds, policy definitions, and applicability decisions, are treated as externally generated and frozen before governance execution. Their predictive validity is evaluated independently from the blockchain layer.

The adversary is assumed to have access to the public contract interface and may construct or replay transactions using manipulated governance inputs. The adversary may also control an account that lacks the role required for a requested operation, or may attempt to misuse a legitimately authorised operational account outside the decision constraints already recorded by the governance system. The considered attacks do not require modification of the deployed contract or compromise of the underlying ledger.

Within this model, the adversary may attempt the following classes of invalid operation: (i) invoke privileged case-submission, decision, review, or execution functions without the required role; (ii) substitute a model or policy authority that is inconsistent with the registered model--policy binding; (iii) alter a submitted score, risk state, or action so that the claimed decision is inconsistent with the authorised policy; (iv) execute an action different from the authorised final action; (v) replay execution, review-resolution, or finalisation operations after the corresponding lifecycle transition has already occurred; (vi) introduce an unregistered or incorrect automated-refiner authority; (vii) initiate governed operations while the system is administratively paused; or (viii) attempt to use an authority that has been revoked.

The security objective is fail-closed decision governance: an invalid operation should be rejected before it can cross the corresponding governance boundary, whereas a legitimate operation satisfying all applicable role, authority, policy, and lifecycle constraints should remain executable. Rejection concerns governance validity rather than predictive correctness. A successfully governed decision should not be interpreted as cryptographic attestation that the upstream ML or LLM inference itself was executed correctly.

Attacks against the consensus protocol, compromise of the authority-administration trust root, extraction of private keys, denial-of-service at the underlying blockchain-network layer, and cryptographic attestation of model execution are outside the scope of the present prototype. Likewise, the evaluation does not claim production-network throughput or live-operator fraud validation.

\subsection{Security Objectives}
\label{subsec:security-objectives}
The threat model above leads to eight security objectives for the governance mechanism. Let a governed case be associated with a dataset authority $D$, model authority $M$, policy authority $P$, optional refiner authority $R$, claimed decision state $s$, authorised final action $a$, and lifecycle state $\ell$. The governance layer should satisfy the following objectives.

\textbf{O1---Role authorisation.}
A state-changing governance operation shall be accepted only when the calling account holds the role required for that operation. Case submission, initial-decision submission, review resolution, execution, and authority administration remain separated by explicit operational privileges.

\textbf{O2---Authority-binding integrity.}
A governed decision shall reference registered and active authorities whose bindings are mutually consistent. In particular, the submitted model and policy must correspond to the dataset authority of the case, and the policy must be registered for the intended model and case context. Substitution of an otherwise active but incorrectly bound model or policy must fail closed.

\textbf{O3---Decision consistency.}
The decision state submitted to governance shall be consistent with the frozen score-to-state rule encoded by the authorised policy, and the corresponding decision action shall preserve the policy-defined state semantics. Governance must reject a caller-supplied state or action that contradicts the authorised decision mapping.

\textbf{O4---Action integrity.}
Once a final action has been authorised for a case, the execution layer shall permit only that action. An executor may not replace the authorised action with an alternative action, even when the executor itself holds the correct execution role.

\textbf{O5---Lifecycle and replay safety.}
Every state-changing operation shall occur only from a lifecycle state in which that transition is permitted. Operations that have already been consumed, including execution, review resolution, and finalisation, shall not be accepted a second time. A terminal case shall not be reopened through replay of an earlier transition.

\textbf{O6---Refiner-authority integrity.}
Automated refinement shall be permitted only when the active policy explicitly authorises automated refinement and the submitted refiner authority matches the registered, applicable refiner authority. A fabricated, unregistered, or incorrectly bound refiner must not be able to resolve a review case.

\textbf{O7---Administrative pause safety.}
When governance is administratively paused, operational state-changing functions covered by the pause mechanism shall fail closed. Possession of an otherwise valid operational role must not bypass the paused state.

\textbf{O8---Authority-revocation safety.}
Revocation shall prevent a previously valid authority from authorising subsequent governed operations. In particular, an authority that is no longer active must not be usable to create or execute a new governed decision transition, while already-finalised historical records remain auditable.

These objectives are intentionally separable. This enables the evaluation to test the complete governance mechanism and then remove one enforcement capability at a time to determine whether the corresponding invalid operations become admissible while unrelated protections remain intact.

\section{Authority-Bound Governance Architecture}
\label{sec:governance}

\subsection{Governance Flow}
\label{subsec:governance-flow}
\begin{figure}[t]
\centering
\includegraphics[
    width=0.93\textwidth,
    trim=2.5cm 8.0cm 0.0cm 2.0cm,
    clip
]{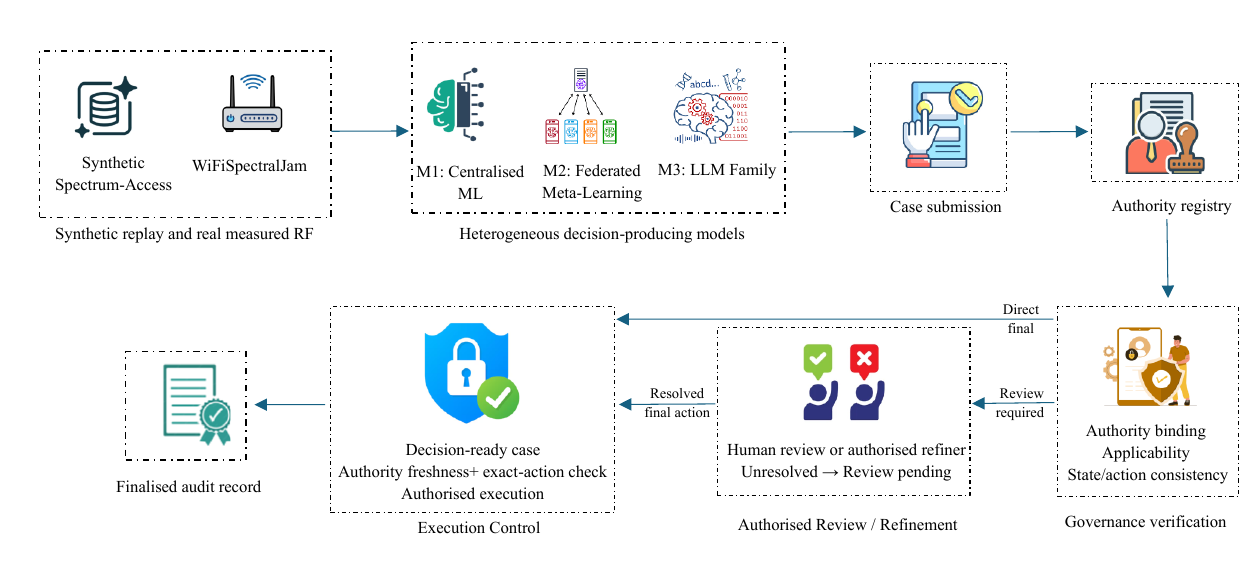}
\caption{Overview of the authority-bound governance framework. Frozen decision evidence from the controlled synthetic spectrum-access branch and the independently measured WiFiSpectralJam RF branch is submitted to a shared governance path. Registered authorities constrain binding, applicability, and state--action consistency. Direct final decisions proceed to execution control, whereas review-required cases require human review or an authorised refiner; unresolved cases remain in review pending. Before execution, decision authorities are rechecked for freshness and the submitted action must match the authorised final action. Executed cases are subsequently finalised as auditable governance records.}
\label{fig:framework-overview}
\end{figure}

The governance architecture operates as a downstream control plane between frozen AI-generated security evidence and operational execution. It does not reproduce the upstream scientific pipeline. Its role is to determine whether a submitted decision is admissible under the authorities and decision rules registered for that deployment. The flow has five stages: case registration, authority-bound decision validation, optional review, authorised execution, and finalisation. Figure~\ref{fig:framework-overview} provides an overview of this workflow and shows how heterogeneous upstream decision sources enter the shared governance path while direct, review-required, and unresolved-review cases retain their respective lifecycle outcomes.

A governed case is created by a role-authorised submitter and is bound at creation time to a case type, dataset authority, and non-zero case-evidence hash. The contract accepts the case only when the referenced dataset authority is registered, active, and applicable to the submitted case type. This establishes the governance context before any model-generated decision is considered.

An initial decision is submitted by a validator together with the model authority, policy authority, decision evidence, score representation, claimed state, and claimed action. The governance layer first verifies that the dataset, model, and policy authorities are active and mutually compatible. In particular, the model and policy must refer to the dataset authority already bound to the case, the policy must be bound to the submitted model, and the policy must support the case type. Only after these authority checks does the contract evaluate the submitted decision against the registered policy semantics.

For policies that expose a numerical decision score, the score is represented using a fixed-point integer encoding and is evaluated against the policy boundaries registered on-chain. The resulting expected state and action are recomputed by the governance logic and compared with the caller-supplied state and action. A mismatch causes the transaction to revert rather than recording an inconsistent decision. If the validated action is already final, the case enters the decision-ready state. If the validated action requires review, the case instead enters the review pending state and no final operational action is authorised at that point.

Review resolution is itself a governed transition. A reviewer may resolve only a case currently in the review pending state. Human review and automated refinement are represented as distinct resolution modes. Automated refinement requires explicit policy-level authorisation and an active refiner authority that is bound to the relevant dataset, model, and case type. After a valid review resolution, the selected final action becomes authorised and the case moves to the decision-ready state.

Execution is separated from decision formation. An executor may act only on a decision-ready case and must submit exactly the action already authorised by governance. Before execution, the contract rechecks the current activity of the decision authorities so that subsequent revocation can prevent a stale authority from enabling a new operational action. Once execution succeeds, the case enters the executed state and may subsequently be finalised as an immutable historical governance record.

\subsection{Authority Model}
\label{subsec:authority-model}
The governance mechanism represents decision provenance through explicit on-chain authorities rather than through mutable textual identifiers. Each authority is referenced by a non-zero cryptographic hash corresponding to a frozen upstream artefact or authority definition. Registration establishes which dataset, model, policy, and optional refinement component may participate in a governed decision, while later decision transactions consume these registered references rather than redefining them.

\textbf{Dataset authority.}
The dataset authority defines the admissible data context for a governed case and the case types that the corresponding evidence is permitted to support. A dataset authority must be registered before a case or dependent model authority can use it. Registration records the authority hash, supported-case-type mask, activation state, registration time, and eventual revocation time. A case can subsequently be created only against an active dataset authority that supports its declared case type.

\textbf{Model authority.}
A model authority is registered against one dataset authority and a supported-case-type mask. Registration is accepted only when the parent dataset authority is already active and the model's supported case types are a subset of those authorised by that dataset. A model authority cannot later be treated as a dataset-independent identifier because its admissible scope is established at registration.

\textbf{Policy authority.}
A policy authority binds the operational decision rule to a specific dataset, model, and case type. For a four-state policy, it additionally stores the frozen decision boundaries used for score-to-state mapping; policy metadata also records whether a hard gate is enabled and whether automated refinement is authorised. Policy registration verifies that the referenced dataset and model authorities are active, that the model belongs to the stated dataset, and that both authorities support the intended case type. These checks prevent an otherwise valid policy identifier from being transplanted into a different model or dataset context.

\textbf{Refiner authority.}
Automated refinement is represented through a separate component authority rather than being implicitly granted to the primary model. The current governed component type is a refiner, whose authority is bound to a dataset authority, model authority, and supported-case-type set. If a policy authorises automated refinement, the policy must reference an active compatible refiner authority and non-zero refinement-applicability evidence. If automated refinement is not authorised, the policy must not carry a refiner authority hash. This distinction preserves cases in which refinement is scientifically unsupported rather than manufacturing a downstream refinement capability.

Authorities are revocable. Revocation changes the authority from active to inactive and records the revocation time without rewriting the authority's historical identity. Subsequent state-changing operations that require the revoked authority fail the active-authority checks, while certificates produced before revocation remain associated with the original authority hashes for auditability.

Finally, each contract instance is bound to a non-zero immutable deployment identifier and a constructor-supplied experiment tag, while the contract version is fixed as governance-2. The deployment identifier separates otherwise structurally similar governance instances and is included in generated decision certificates. This allows the same authority architecture to be instantiated across heterogeneous experimental branches without treating decisions originating from different deployments as interchangeable.

\subsection{Decision Semantics and Applicability}
\label{subsec:decision-semantics}
The governance contract supports a shared state vocabulary while preserving case-type-specific action semantics. This distinction is necessary because the two empirical branches do not represent the same security task. A spectrum-access request concerns an access-control decision, whereas an RF-interference event concerns a measured RF condition. The governance layer reuses common uncertainty states without forcing the two domains to share an artificial operational meaning.

Two policy modes are supported. Under four-state, a numerical score must be available and three ordered policy boundaries, $b_1<b_2<b_3$, partition the score into four states:
\begin{equation}
s(p)=
\begin{cases}
\plainid{NO}, & p<b_1,\\
\plainid{MAYBE-LOW}, & b_1\leq p<b_2,\\
\plainid{MAYBE-HIGH}, & b_2\leq p<b_3,\\
\plainid{YES}, & p\geq b_3.
\end{cases}
\label{eq:score-state}
\end{equation}
The contract recomputes this state from the submitted score and registered policy boundaries; the caller cannot independently assign the state. Both intermediate states are review states. Their distinction preserves the upstream ordering of uncertainty/risk, but neither is treated as a directly executable final action.

The state-to-action mapping is case-type aware. For both case types, \plainid{NO} maps to \plainid{ALLOW}, while \plainid{MAYBE-LOW} and \plainid{MAYBE-HIGH} map to \plainid{REVIEW}. A \plainid{YES} state maps to \plainid{DENY} for a spectrum-access request but to \plainid{MITIGATE} for an RF-interference event. This allows the contract to use the same high-level state vocabulary without treating RF interference as telecom fraud or forcing an RF mitigation event into an access-denial interpretation.

The synthetic spectrum-access branch additionally supports a deterministic hard-gate condition. When the upstream evidence indicates that the registered hard gate has been triggered, the contract accepts hard-fraud only for a spectrum-access request, only when the active policy explicitly enables that gate, and only when non-zero hard-gate evidence is supplied. hard-fraud maps directly to \plainid{DENY}. This state is deliberately unavailable to the real-RF branch: policies registered for RF-interference event are prohibited from enabling the hard gate.

The second policy mode, review-all, represents a case for which no admissible automated four-state decision is available. Its policy boundaries must remain zero, and the governance contract assigns the state unavailable, which maps to \plainid{REVIEW}. This is a fail-closed representation of non-applicability rather than an artificial prediction. In the real-RF evaluation, for example, the frozen M3-Base stream uses review-all because its validation probability was not directionally informative; the governance layer does not reverse the score direction, invent alternative boundaries, or create a synthetic direct decision.

Applicability is treated in the same way for optional downstream capabilities. A hard gate is admitted only for the case type and policy that explicitly support it, and automated refinement is admitted only when a compatible refiner has been authorised. When these capabilities are unsupported, their absence is preserved rather than replaced by a fabricated equivalent. The governance plane enforces the semantics supplied by the frozen upstream authorities while preventing unsupported functionality from being introduced during downstream execution.

\subsection{Governed Lifecycle and Decision Certificates}
\label{subsec:lifecycle-certificates}

Each governed case follows an explicit state machine rather than a sequence of unconstrained logging operations. The principal lifecycle states are submitted, review pending, decision-ready, executed, and finalised, with cancelled providing an administrative terminal path for cases that have not yet been executed. The admissible transition depends on both the current lifecycle state and the role invoking the next operation.

A newly registered case enters submitted. Exactly one initial governed decision may then be accepted. If the policy-validated initial action is final, the case moves directly to decision-ready. If the validated action is \plainid{REVIEW}, the case instead enters review pending, and execution remains unavailable until an admissible review resolution has established a final action. A successful review resolution moves the case to decision-ready. From this state, an authorised executor may execute only the previously authorised final action, after which the case enters executed. Finalisation is permitted only after successful execution and moves the case to the terminal finalised state.

The lifecycle enforces single-use transitions. An initial decision cannot be submitted twice, a review cannot be resolved after it has already been consumed, an executed case cannot be executed a second time, and a finalised case cannot be finalised again. Terminal finalised and cancelled cases are excluded from subsequent operational transitions. This state-dependent enforcement provides replay protection at the governance-operation level rather than relying solely on transaction identifiers or an off-chain duplicate check.

An authority administrator may cancel a non-executed case by supplying a non-zero cancellation-reason hash. Cancellation is prohibited after execution and is itself terminal. This path allows a pending case to be withdrawn without fabricating an execution outcome or modifying an already executed historical decision.

Accepted decision transitions produce decision certificates. An initial-decision certificate binds the case and deployment identifiers to the case evidence, dataset authority, model authority, policy authority, submitted decision evidence, policy boundaries, resulting risk state and action, applicability-related evidence, actor, and timestamp. If review is subsequently resolved, a review-resolution certificate records the resolution mode, review evidence, optional refiner authority, and final resolved action while retaining the relevant authority and decision context from the initial certificate.

Certificates within a reviewed case are linked through a parent-certificate hash. The initial decision has no parent, whereas the review-resolution certificate references the latest preceding certificate. Each certificate digest is computed after verifying that its deployment identifier matches the active contract deployment and is additionally scoped to the blockchain chain identifier and the deployed contract address. This prevents an otherwise identical certificate payload from being interpreted as belonging to a different contract instance or chain.

Execution and finalisation are deliberately represented separately from decision certificates. Execution records the authorised executor and a non-zero execution-evidence hash and emits an execution event; finalisation then seals the already executed case. The certificate chain records how the governed decision was formed and, where relevant, reviewed. The case lifecycle and execution evidence show whether the authorised decision was later executed. The design does not interpret these records as proof that the upstream AI computation itself occurred correctly.

Algorithm~\ref{alg:authority-bound-governance} summarises the governance
procedure from case admission to finalisation, showing how frozen decision
evidence progresses through the registered authority, policy, review, and
execution lifecycle.

\begin{algorithm}[H]
\caption{Authority-Bound Governance of an AI Security Decision}
\label{alg:authority-bound-governance}
\footnotesize
\begin{algorithmic}[1]

\Require Case $c$, evidence $e$, dataset authority $D$, model authority $M$,
policy authority $P$, optional refiner authority $R$, submitted score $p$,
claimed state $s$, and claimed action $a$
\Ensure Governed lifecycle state and, where applicable, an authorised final action

\State Verify that the caller holds the role required for the requested transition
\State Verify that $c$ carries nonzero evidence $e$ and that $D$ is registered, active, and applicable to the case type
\State Verify that $M$ is active and bound to $D$ and the case type
\State Verify that $P$ is active and compatible with $D$, $M$, and the case type
\State Derive $(\hat{s},\hat{a})$ under $P$ using the applicable four-state,
review-all, or authorised hard-gate semantics

\If{$s \neq \hat{s}$ or $a \neq \hat{a}$}
    \State \textbf{reject}
\EndIf

\If{$\hat{a}$ requires review}
    \State Move the case to review pending
    \If{a human review resolution is submitted}
        \State Verify the authorised reviewer and required review evidence
        \State Resolve the authorised final action
        \State Move the case to decision-ready
    \ElsIf{an automated review resolution is submitted}
        \State Verify that automated review is permitted by $P$
        \State Verify that $R$ is active, compatible, and supported by nonzero applicability evidence
        \State Resolve the authorised final action under the registered review policy
        \State Move the case to decision-ready
    \Else
        \State \Return review pending
    \EndIf
\Else
    \State Record the authorised final action
    \State Move the case to decision-ready
\EndIf

\State Recheck the active decision authorities required for execution
\State Require the submitted execution action to equal the authorised final action
\State Require that the case has not already been executed
\State Execute the authorised action and move the case to executed
\State Finalise the executed case and seal its governance record
\State \Return finalised

\end{algorithmic}
\end{algorithm}

The execution-stage authority check is deliberately repeated rather than
inferred from successful decision admission. An authority may become
inactive after the initial decision has been accepted but before execution.
The exact-action check similarly prevents an executor from substituting a
different operational action for the one authorised by the governed
decision. Finalisation seals an already executed lifecycle record and does
not constitute a new scientific or operational authorisation.

\subsection{Role Separation and Administrative Controls}
\label{subsec:roles-admin}
The contract separates operational responsibilities through five explicit roles: case submitter, validator, reviewer, executor, and authority administrator. A role is granted to an account by the contract owner and is checked at the entry point of each protected operation. Holding one role does not automatically grant another.

The case submitter creates a governed case but does not determine its decision. The validator submits the initial AI-derived decision and its supporting authority references. Cases routed to review can be resolved only by a reviewer, while execution and finalisation are restricted to an executor. Authority administrators register and revoke dataset, model, policy, and component authorities and may cancel eligible non-executed cases. This separation prevents a generic authorised account from exercising every stage of the decision lifecycle.

Administrative ownership is kept distinct from operational authority. The owner controls role assignment, pause state, and ownership transfer. Ownership alone does not satisfy the role checks used for case submission, validation, review, or execution. An owner that needs to perform one of these operations must be granted the corresponding operational role explicitly. The owner is, however, accepted by the authority-administration check and can manage the authority registry directly.

The pause mechanism provides an emergency control over governed state changes. Operations protected by the pause guard revert while the contract is paused, even if the caller otherwise holds the required role. Authority registration and revocation are covered by the same guard. Ownership transfer also clears the previous owner's role mask, so operational privileges held by the former owner do not survive the transfer implicitly.

\section{Security Properties and Enforcement Mechanisms}
\label{sec:security-properties}

\subsection{Enforcement Model}
\label{subsec:enforcement-model}
The security objectives in Section~\ref{subsec:security-objectives} are enforced as admissibility conditions on state-changing governance operations. Let $x$ denote a requested governance operation on case $c$, and let $\mathcal{G}(x,c)$ denote the set of checks that apply to that operation. Depending on the operation and case state, these checks cover role authorisation, active authority status, authority compatibility, decision consistency, final-action integrity, lifecycle validity, refiner applicability, and the administrative pause state. The contract accepts the operation only when every applicable check succeeds:
\begin{equation}
\mathrm{Accept}(x,c)=1
\iff
\bigwedge_{g \in \mathcal{G}(x,c)} g(x,c)=1 .
\label{eq:governance-acceptance}
\end{equation}

The set $\mathcal{G}(x,c)$ is operation-specific. A case submission, for example, does not yet have a model or policy binding to verify, whereas an initial decision requires both authority binding and decision-consistency checks. Review resolution introduces additional applicability checks when automated refinement is requested. Execution requires a final authorised action and rechecks the activity of the authorities on which that decision depends.

The implementation follows a fail-closed rule. If any applicable predicate in \eqref{eq:governance-acceptance} fails, the corresponding state-changing call reverts and the attempted governance transition is not committed. Passing the checks establishes governance admissibility under the registered authorities and contract rules. Table~\ref{tab:security-enforcement-map} maps each security objective to its principal enforcement mechanism and the corresponding adversarial probes.

\begingroup
\footnotesize
\renewcommand{\arraystretch}{1.12}
\setlength{\tabcolsep}{3pt}
\setlength\LTleft{0pt}
\setlength\LTright{0pt}

\begin{longtable}{@{}
    >{\centering\arraybackslash}p{0.07\linewidth}
    >{\raggedright\arraybackslash}p{0.18\linewidth}
    >{\raggedright\arraybackslash}p{0.38\linewidth}
    >{\raggedright\arraybackslash}p{\dimexpr0.37\linewidth-6\tabcolsep\relax}
@{}}
	\caption{Security objectives, enforcement mechanisms, and adversarial probes.}
	\label{tab:security-enforcement-map}\\
	
	\toprule
	\textbf{Obj.} &
	\textbf{Property} &
	\textbf{Principal enforcement mechanism} &
	\textbf{Representative adversarial probe(s)} \\
	\midrule
	\endfirsthead
	
	\multicolumn{4}{c}{\tablename\ \thetable\ -- continued from previous page} \\
	\toprule
	\textbf{Obj.} &
	\textbf{Property} &
	\textbf{Principal enforcement mechanism} &
	\textbf{Representative adversarial probe(s)} \\
	\midrule
	\endhead
	
	\midrule
	\multicolumn{4}{r}{Continued on next page} \\
	\endfoot
	
	\bottomrule
	\endlastfoot
	
	O1 &
	Role authorisation &
	Role checks restrict case submission, initial decision, review resolution, execution, and authority administration to the corresponding authorised actors. &
	Unauthorised case submission, decision submission, execution, or review resolution (A01, A02, A06, A10). \\
	
	O2 &
	Authority-binding integrity &
	Active dataset, model, and policy authorities are checked for dataset, model--policy, and case-type compatibility before a governed decision is accepted. &
	Substitution of an active but incorrectly bound model or policy (A03, A04). \\
	
	O3 &
	Decision consistency &
	The contract derives the expected state from the submitted score and registered boundaries, then derives the expected action from the state and case type. Caller-supplied values must match. &
	Inconsistent submitted score, state, or action (A05). \\
	
	O4 &
	Action integrity &
	Execution is restricted to the final action already authorised for the case. &
	Authorised executor submits a different final action (A07). \\
	
	O5 &
	Lifecycle and replay safety &
	Lifecycle guards, single-use flags, and terminal-state checks restrict each operation to its admissible state and reject repeated transitions. &
	Repeated execution, finalisation, or review resolution (A08, A09, A12). \\
	
	O6 &
	Refiner-authority integrity &
	Automated refinement requires policy-level authorisation and a registered, active refiner bound to the relevant dataset, model, and case type. &
	Unregistered or tampered refiner authority (A11). \\
	
	O7 &
	Administrative pause safety &
	State-changing operations covered by the pause guard revert while governance is paused. &
	Authorised operation attempted during pause (A13). \\
	
	O8 &
	Authority-revocation safety &
	Operations that depend on an authority require them to remain active; decision authorities are checked again before execution. &
	Use of a revoked dataset authority (A14). \\
	
\end{longtable}
\endgroup

\subsection{Property-Level Security Analysis}
\label{subsec:property-analysis}
We use the enforcement model to construct a property-level security argument under the trust assumptions stated in Section~\ref{subsec:threat-model}. Each property below follows from explicit contract guards and lifecycle conditions and is subsequently exercised by the adversarial, stateful-invariant, and capability-ablation tests. These properties concern the governance boundary and do not constitute claims about the predictive correctness of the upstream AI model or the security of the underlying blockchain consensus protocol.

\textbf{Authority-consistent decision admission.}
A decision cannot be admitted solely because its model and policy hashes refer to registered objects. The dataset authority must be active, the model and policy must also be active, both must refer to the dataset already bound to the case, and the policy must be registered for the submitted model and case type. An active authority taken from another admissible context is rejected when these bindings do not match. This limits cross-context substitution of otherwise valid governance artefacts.

\textbf{Decision and execution integrity.}
For a four-state policy, the contract derives the expected risk state from the submitted score and the registered policy boundaries, then derives the expected action from that state and the case type. The submitted state and action must match these values. Once a final action has been authorised, execution performs a separate equality check against that stored action. A caller with a valid executor role cannot use that role to replace the authorised outcome with another executable action.

\textbf{Lifecycle and replay protection.}
Governance operations are tied to specific lifecycle states. An initial decision is accepted only from submitted, review resolution only from review pending, execution only from decision-ready, and finalisation only after execution. Single-use flags and terminal-state checks reject repeated decision, review, execution, and finalisation attempts. Replay protection is thus attached to the governed state transition itself rather than to an off-chain assumption that a request will be submitted only once.

\textbf{Authority freshness at operational use.}
Registration establishes that an authority is valid at the time it enters the governance registry, but later state-changing operations still require the relevant authority to remain active. The contract also rechecks the authorities supporting a decision immediately before execution. A decision that was admissible earlier cannot be used for a new execution after one of its required authorities has been revoked. Finalisation is treated differently: it seals an action that has already been executed and does not perform a new active-authority check.

\textbf{Applicability-preserving refinement.}
Automated refinement is not inferred from the presence of a review state. It must be enabled by the active policy and backed by a registered refiner authority with compatible dataset, model, and case-type bindings. When automated refinement is not authorised, the refiner hash must remain absent. This prevents the governance layer from creating an unsupported second-stage decision capability merely because a case requires review.

\textbf{Deployment-scoped decision certificates.}
Decision certificates carry the deployment identifier and are hashed together with the blockchain chain identifier and the deployed contract address. The same certificate payload presented in another contract instance or on another chain would not retain the same certificate digest. This provides deployment-scoped provenance for accepted decisions. It should not be interpreted as a general proof that the underlying off-chain evidence could never be resubmitted elsewhere; the property concerns the identity and provenance of the governance certificate itself.

These properties define the security boundary evaluated experimentally in Section~\ref{sec:experiments}. The capability-ablation study removes one enforcement function at a time and tests whether the invalid operations associated with that function become admissible while unrelated protections remain active.

\section{Experimental Methodology}
\label{sec:experiments}

\subsection{Evaluation Design}
\label{subsec:evaluation-design}

The evaluation uses two independent security-decision branches followed by the same governance mechanism. The first branch is a controlled synthetic spectrum-access setting with four heterogeneous upstream decision configurations. The second uses WiFiSpectralJam, a real measured-RF dataset, to test the governance design under a different data source and operational decision meaning. The real-RF branch is not presented as validation on live telecom fraud traffic; it provides an independent measured-data setting in which the same authority, applicability, lifecycle, and execution rules can be exercised.

The scientific decision pipelines and the governance evaluation are kept separate. Dataset definitions, model outputs, decision boundaries, policy mappings, and refinement applicability are fixed upstream before blockchain execution. The governance experiments consume these artefacts without retraining models, recalibrating scores, selecting new thresholds, or changing state--action mappings. This separation allows predictive behaviour and governance behaviour to be evaluated as different parts of the system.

The experiments address three questions. First, we examine whether heterogeneous frozen decision streams can be processed under a common governance mechanism without changing their native semantics. Second, we test whether invalid role, authority, decision, lifecycle, refinement, pause, and revocation operations are rejected by the full governance contract. Third, we remove individual enforcement capabilities and measure which tested invalid operations become admissible, while checking that legitimate execution paths remain available.

\subsection{Controlled Synthetic Spectrum-Access Branch}
\label{subsec:synthetic-branch}

The synthetic branch uses a controlled post-training replay of 100,000 spectrum-access requests. It was retained as a reproducible environment in which the complete decision lifecycle, including deterministic hard-gate cases, uncertain states, optional refinement, final actions, and blockchain execution, can be exercised at scale. The data should not be interpreted as traffic collected from a live telecom operator.

Four frozen upstream configurations are evaluated: M1, M2, M3-Base, and M3-QLoRA. M1 uses a centralised machine-learning ensemble risk signal, M2 uses a federated meta-model over retained detector outputs, and the M3 family provides two LLM-based risk sources: a structured base-model variant and a Quantized Low-Rank Adaptation (QLoRA)-tuned sequence-classification variant. These configurations are treated as heterogeneous upstream decision sources rather than as the main contribution of this paper. Their scores and retained decision artefacts are consumed by the governance layer after the scientific pipeline has been completed.

All four configurations are evaluated under the frozen official recall-first operating policy. The policy maps non-hard scores into the \plainid{NO}, \plainid{MAYBE-LOW}, \plainid{MAYBE-HIGH}, and \plainid{YES} states using configuration-specific frozen boundaries. The synthetic branch also contains a deterministic hard-gate path. Requests satisfying that upstream condition are assigned hard-fraud and do not rely on the non-hard score for their state assignment. In the 100,000-request replay, 4,500 requests follow this hard-gate path.

The governance input for each request is constructed from the already-fixed scientific evidence: case type, dataset authority, model authority, policy authority, decision evidence, score representation where applicable, risk state, requested action, and any authorised refinement evidence. The contract does not derive a new scientific label from the synthetic data. Its task begins after these upstream artefacts exist and consists of checking whether the submitted evidence is authorised and internally consistent before execution.

The same 100,000-request population is replayed for each upstream configuration. This keeps the request population fixed while the source of the non-hard decision evidence changes. Governance outcomes can then be compared without treating differences among M1, M2, M3-Base, and M3-QLoRA as differences in the blockchain mechanism itself.

\subsection{WiFiSpectralJam Dataset and Data Preparation}
\label{subsec:realrf-data}
The real-RF branch uses WiFiSpectralJam, a public commodity Wi-Fi spectral-scan dataset collected under benign operation and controlled RF-jamming conditions \cite{herzalla2026wifispectraljam}. The release contains 96,090 scan files and 522,771,130 spectral observations acquired with a Raspberry Pi Compute Module 4 and a Qualcomm Atheros QCA9880 Wi-Fi interface. Controlled jamming was generated with a HackRF One, while benign data include both RF-chamber floor measurements and real-world indoor background captures. Active and passive scans cover the 2.4 and 5~GHz bands. The malicious class denotes a jammer-present spectral condition; it does not represent packet-level service degradation or telecom fraud.

We use the released file-level aggregate representation rather than individual spectral rows. File-level processing is important because observations within the same capture share the same acquisition condition and should not be split across training and test partitions. The source representation contains 96,090 rows and 108 columns. Following the canonical benchmark representation, 63 columns are used as model predictors. These features are formed from seven measured quantities---average power, base power, maximum magnitude, noise, relative power, received signal strength indicator (RSSI), and total gain---with nine summary statistics retained for each quantity: mean, standard deviation, minimum, first quartile, median, third quartile, maximum, skewness, and kurtosis.

The remaining 45 columns are excluded from model inputs. They contain identifiers, the target label, acquisition-condition metadata, quality-control or size information, and variables outside the canonical benchmark feature set. Examples include file paths, benign subtype, location, waveform, jammer power, target channel, collection environment, row-count and file-size information, centre-frequency summaries, raw unconverted RSSI summaries, and unique-count variables. Scan mode and frequency band are retained only as acquisition-domain context for split auditing and subgroup construction; neither is used as a predictor. No synthetic predictors are added to the real-RF representation.

The data are separated at file level. We follow the matched-distribution outer-test logic of the released technical-validation benchmark, using seed 1 and a 20\% final-test partition. The remaining 80\% development population is split again, with 20\% of that population reserved for validation. This gives 61,497 training files, 15,375 validation files, and 19,218 final-test files, corresponding to approximately 64\%, 16\%, and 20\% of the full dataset. The development split is stratified by class label and acquisition domain, where the latter distinguishes active 5~GHz, passive 2.4~GHz, and passive 5~GHz captures.

Preprocessing is fitted using the training partition only. Missing predictor values are handled with median imputation, and a standardisation transform is fitted on the same training data for models that require scaled inputs. The fitted preprocessing objects are then reused for validation and final-test inference. The final-test predictor matrix and labels remain unopened while model development and policy construction are performed. They are materialised only after the upstream model and policy artefacts have been frozen.

\subsection{Real-RF Models and Frozen Decision Streams}
\label{subsec:realrf-models}
The real-RF predictive evaluation is anchored to the classifier families reported with WiFiSpectralJam. We evaluate Logistic Regression (LR), Random Forest, HistGradientBoosting (HistGB), XGBoost (XGB), LightGBM (LGBM), and a multilayer perceptron (MLP) using the canonical 63-feature representation described in Section~\ref{subsec:realrf-data}. These classifier families provide a benchmark-aligned comparison with the model families reported for WiFiSpectralJam
\cite{herzalla2026wifispectraljam}; the train/validation/final-test protocol, frozen seeds, and retained outputs used here remain specific to the present evaluation. The decision threshold for the binary benchmark comparison is fixed at 0.5. Models are evaluated for seeds 1, 7, and 42, giving 18 baseline final-test streams.

The baseline configurations follow the released WiFiSpectralJam benchmark. Logistic Regression uses median imputation followed by standardisation and a maximum of 3,000 iterations. Random Forest uses 300 trees with balanced-subsample class weighting. HistGradientBoosting uses the corresponding library defaults with a seed-specific random state. XGBoost uses 300 estimators, maximum depth 6, learning rate 0.05, subsample and column-subsample rates of 0.9, and histogram-based tree construction. LightGBM uses 300 estimators, learning rate 0.05, and balanced class weighting. The MLP has hidden layers of 256, 128, and 64 units, rectified linear unit (ReLU) activations, batch normalisation after the first two hidden layers, dropout rates of 0.30, 0.25, and 0.20, and Adam optimisation with a learning rate of $10^{-3}$. It is trained for at most 25 epochs with early stopping based on an internal training-only validation split.

Four of these benchmark models---Random Forest, XGBoost, LightGBM, and MLP---already exist as identically configured members of the frozen M1 pipeline and are reused rather than fitted again. Logistic Regression and HistGradientBoosting are fitted separately for the three fixed seeds. No model is selected by final-test performance, and no best-seed selection is performed.

The benchmark comparison is treated as a same-dataset, benchmark-aligned reference rather than an exact reproduction of the published WiFiSpectralJam results. The published matched-distribution benchmark reports mean performance across seeds 1, 7, and 42. Our protocol fixes the official-style 20\% outer test using seed 1, then evaluates the three model seeds on that same sealed final-test population. We also retain a separate validation partition within the development data and do not refit models on the combined training and validation sets before final testing. The comparison is used to check whether the predictive pipeline remains consistent with the published benchmark under these protocol differences; the numerical comparison is reported in Section~\ref{sec:results}.

The governance evaluation uses a separate set of eight frozen policy streams. M1 and M2 each contribute three streams corresponding to seeds 1, 7, and 42. M3-QLoRA contributes one frozen stream, while M3-Base contributes one review-all stream. M1 represents the centralised ensemble branch, M2 is a federated meta-model constructed over retained M1 detector outputs, and the M3 variants provide the LLM-family decision sources.

For M1, M2, and M3-QLoRA, the four-state policy boundaries are derived from development data and frozen before the final test is opened. M3-Base does not receive an alternative threshold or reversed score direction when its validation probability is not directionally informative. Its frozen policy is review-all, so final-test cases are assigned unavailable and routed to review. No hard-fraud state is used in the real-RF branch, and no automated second-stage refiner is authorised.

The final-test outputs of these eight streams are converted into governed RF-interference event cases using the frozen state--action mapping defined in Section~\ref{subsec:decision-semantics}. Model scores and policy states are produced before blockchain execution. Governance-2 consumes those frozen outputs without retraining a model, changing a boundary, reversing a score, or resolving review cases on behalf of the upstream scientific pipeline. Table~\ref{tab:realrf-baselines} summarises the six benchmark classifier families and their principal configurations.

\begin{table}[t]
\caption{WiFiSpectralJam benchmark models used for the real-RF reference comparison.}
\label{tab:realrf-baselines}
\centering
\small
\renewcommand{\arraystretch}{1.10}

\begin{tabular*}{\columnwidth}{@{\extracolsep{\fill}}lll@{}}
\toprule
\textbf{Model} &
\textbf{Preprocessing} &
\textbf{Principal configuration} \\
\midrule

LR &
Impute + scale &
Maximum 3,000 iterations \\

Random Forest &
Impute &
300 trees; balanced-subsample weighting \\

HistGB &
Impute &
Library defaults; seed-specific random state \\

XGB &
Impute &
300 trees; depth 6; learning rate 0.05; subsample 0.9 \\

LGBM &
Impute &
300 trees; learning rate 0.05; balanced class weighting \\

MLP &
Impute + scale &
256--128--64; Adam; dropout; early stopping \\

\bottomrule
\end{tabular*}
\end{table}

\subsection{Governance Execution and Security Tests}
\label{subsec:governance-evaluation}
Governance execution begins only after the upstream scientific artefacts for a decision stream have been frozen. Each deployment receives the corresponding dataset, model, policy, and, where applicable, refiner authorities before case processing starts. Operational roles are assigned to separate accounts for case submission, validation, review, execution, and authority administration. The blockchain layer then receives the frozen case evidence and decision outputs through the lifecycle described in Section~\ref{subsec:lifecycle-certificates}.

The synthetic branch uses the complete 100,000-request replay for each frozen upstream configuration. Cases with a final \plainid{ALLOW} or \plainid{DENY} decision proceed through authorised execution and finalisation. Review cases follow the review path defined by the frozen synthetic policy and may be resolved only when an upstream-authorised review or refinement mechanism is available. Scientific decisions are not recomputed during blockchain execution.

For the real-RF branch, each of the eight frozen policy streams is processed through a separate fresh governance-2 deployment. Each stream contains the same 19,218 final-test captures, giving 153,744 governed stream--case executions. Cases with a direct \plainid{ALLOW} or \plainid{MITIGATE} decision may proceed to execution and finalisation. Cases mapped to \plainid{REVIEW} remain in review pending, because the real-RF experiment contains no scientifically authorised second-stage refiner or review-resolution output. No blockchain transaction is used to manufacture a final decision for these unresolved cases.

Security testing is kept separate from the authoritative governance runs. The tests use fresh isolated contract state and dedicated accounts so that deliberately invalid operations cannot alter the governance records used for the main evaluation. Valid control paths are created first where a particular lifecycle state is needed, after which adversarial inputs are applied at the relevant contract boundary.

The synthetic security test registry contains 14 adversarial probes covering the eight enforcement families defined in Section~\ref{subsec:security-objectives}: role authorisation, authority binding, decision consistency, final-action enforcement, lifecycle and replay protection, refiner authority, administrative pause, and authority revocation. The probes include unauthorised case and decision submission, incompatible model--policy bindings, inconsistent score--state--action claims, unauthorised or incorrect execution, repeated lifecycle operations, invalid refiner use, paused operation, and use of a revoked dataset authority.

The probe registry is objective-driven rather than a sample of arbitrary malformed transactions. Each probe is defined to violate one of the pre-specified governance objectives O1--O8 at a concrete contract boundary, and every objective is represented by at least one negative test. Objectives with multiple independently reachable boundaries, such as role authorisation and lifecycle/replay protection, are represented by multiple probes. The registry is not intended as exhaustive smart-contract verification; its purpose is to exercise the enforcement boundaries defined by the threat model and to support the subsequent capability-ablation analysis.

The real-RF security audit adapts these tests to the RF-specific applicability rules. It includes unauthorised operations, tampered model and policy authorities, inconsistent four-state decisions, incorrect execution actions, replay attempts, unauthorised review resolution, attempted automated refinement when no refiner is authorised, execution or finalisation of unresolved review cases, a false \plainid{NO}/\plainid{ALLOW} claim under review-all, pause enforcement, and authority revocation. These probes exercise both the normal four-state path and the fail-closed review-all path.

A security probe is successful only when the invalid operation is rejected at the intended governance boundary. Legitimate control operations are evaluated separately to ensure that security enforcement does not make the valid lifecycle unusable. The tests assess contract-level governance enforcement; they do not verify the correctness of the upstream model inference or the security of the blockchain consensus protocol.

To broaden the evaluation beyond the fixed adversarial registry, the full governance contract was additionally subjected to randomised stateful invariant testing. Generated transaction sequences exercised valid and invalid role, authority, decision, review, execution, replay, pause, and revocation transitions while checking lifecycle, action, authority-binding, certificate-integrity, and rejected-state immutability invariants. The test used an isolated fresh deployment and did not modify the frozen scientific artefacts.

\subsection{Governance Capability Ablation}
\label{subsec:capability-ablation}
The capability-ablation experiment isolates the contribution of the individual enforcement mechanisms in governance-2. The verified current Solidity source is used as the reference configuration, denoted Full governance. Eight additional contract variants are derived programmatically from that source: Role off, Authority binding off, Decision consistency off, Action enforcement off, Lifecycle/replay off, Refiner authority off, Pause enforcement off, and Authority revocation off. Each variant removes the checks associated with one enforcement family while leaving the remaining governance logic unchanged.

Source derivation is controlled rather than manually implemented as nine independent contracts. The Full governance source is first verified against its expected SHA-256 digest. Each ablated source is then produced through targeted source transformations that remove or neutralise only the checks assigned to the selected capability. Distinct source hashes are required for the eight ablated variants, while Full governance must remain byte-for-byte identical to the verified source.

All nine variants are compiled with Solidity 0.8.24 using the optimiser with 200 runs, intermediate-representation (IR)-based compilation, and the Shanghai Ethereum Virtual Machine (EVM) target. The external application binary interface (ABI) surface is checked after compilation and must remain identical across variants.

The same 14 adversarial probes defined for the synthetic security evaluation are applied to every variant. Each probe is assigned in advance to one of the eight capability families. For a contract variant $v$, the Invalid Acceptance Rate (IAR) is
\begin{equation}
\mathrm{IAR}(v)=
\frac{N_{\mathrm{accepted}}^{\mathrm{invalid}}(v)}
     {N_{\mathrm{probes}}},
\label{eq:iar}
\end{equation}
where $N_{\mathrm{probes}}=14$. A lower IAR indicates that fewer deliberately invalid operations crossed their tested governance boundaries.

We also calculate Security Enforcement Coverage (SEC) at the capability level. A capability is counted as empirically enforced only when all adversarial probes assigned to that capability are rejected:
\begin{equation}
\mathrm{SEC}(v)=
\frac{1}{8}
\sum_{k=1}^{8}
\mathbb{I}\!\left[
\text{all probes in }k\text{ are rejected}
\right].
\label{eq:sec}
\end{equation}
IAR and SEC answer different questions. IAR is probe-weighted, so capabilities represented by several probes contribute more to that rate than capabilities represented by a single probe. SEC gives each of the eight enforcement families equal weight. IAR is not used to rank the security importance of the capabilities.

For each ablated contract, two further checks distinguish the intended effect from unintended weakening elsewhere. The target-probe acceptance rate measures whether probes assigned to the removed capability cross their governance boundary. The non-target protection rate measures whether probes assigned to the other seven capabilities remain rejected. This makes it possible to test whether an ablation exposes its intended boundary without broadly disabling unrelated protections.

Legitimate execution is evaluated separately. A representative M1 control set covers all five synthetic decision states and both valid review outcomes. Duplicate representatives are removed, leaving six distinct valid lifecycle cases. Each case is executed three times for every compiled variant, giving 18 legitimate controls per variant. We report the valid acceptance rate (VAR) and the corresponding false rejection rate (FRR),
\begin{equation}
\mathrm{FRR}(v)=1-\mathrm{VAR}(v).
\label{eq:frr}
\end{equation}
Gas consumption and local transaction latency are recorded from these valid control executions. These measurements characterise prototype contract-execution cost; they are not measurements of production blockchain throughput or consensus performance.

\subsection{Prototype Environment and Metrics}
\label{subsec:prototype-metrics}
The blockchain experiments are conducted on a local Ganache Ethereum-compatible test chain using Web3-based transaction submission and receipt collection. The environment is used to exercise the contract logic under repeatable conditions rather than to emulate a production blockchain network. The capability-ablation contracts are compiled from Solidity 0.8.24, as described in Section~\ref{subsec:capability-ablation}.

For each accepted transaction, the execution receipt provides the gas consumed by the contract call. Client-observed transaction latency is measured from transaction submission until the corresponding receipt is returned. Where a governed case contains several transactions, case-level gas and latency are calculated by summing the measurements for the transactions in that lifecycle. Setup operations---contract deployment, role assignment, and authority registration---are accounted for separately from case-level processing.

Transaction cost is calculated from the observed gas usage and the effective gas price recorded for the transaction. These values describe the local experimental chain and are not interpreted as deployment costs on a public or production network.

Governance correctness is assessed independently of predictive accuracy. For the authoritative executions, we check agreement between frozen upstream states and actions and their on-chain counterparts, expected lifecycle outcomes, authority bindings, certificate integrity, transaction completeness, and the absence of unexpected contract reverts. Review-required real-RF cases are evaluated against their expected review pending state rather than being treated as failed or incomplete direct decisions.

Security evaluation uses the metrics defined in Section~\ref{subsec:capability-ablation}: IAR, SEC, target-probe acceptance, non-target protection, VAR, and FRR. The full contract is evaluated against the adversarial probe set, and the ablation variants are compared with the same probes and legitimate controls. These metrics measure whether the tested governance boundaries accept or reject the operations they are expected to handle.

The real-RF predictive reference comparison is reported separately from the governance metrics. We report average precision (AP), area under the receiver operating characteristic curve (AUROC), and F1 score for the predictive evaluation. F1 score is used to compare the six classifier families with the published WiFiSpectralJam matched-distribution benchmark. This comparison checks the alignment of the upstream predictive pipeline before its frozen outputs are used for governance; it is not a measure of blockchain performance.

For the large real-RF governance execution, we also report transaction count, total gas, recorded confirmation latency, and a latency-based sequential service-rate proxy. If $N_{\mathrm{tx}}$ row-level transactions have recorded confirmation latencies $t_i$ in seconds, the proxy is
\begin{equation}
R_{\mathrm{seq}} =
\frac{N_{\mathrm{tx}}}
     {\sum_{i=1}^{N_{\mathrm{tx}}} t_i}.
\label{eq:service-rate-proxy}
\end{equation}
This quantity describes the rate implied by sequentially summing the observed transaction-confirmation latencies. It is not blockchain network throughput: the experiment does not include a separately measured full-run wall-clock interval, distributed validator network, or consensus-throughput benchmark.

\section{Results}
\label{sec:results}

\subsection{Synthetic Decision Outcomes}
\label{subsec:synthetic-results}
\begin{figure}[!t]
\centering
\includegraphics[width=\columnwidth]
{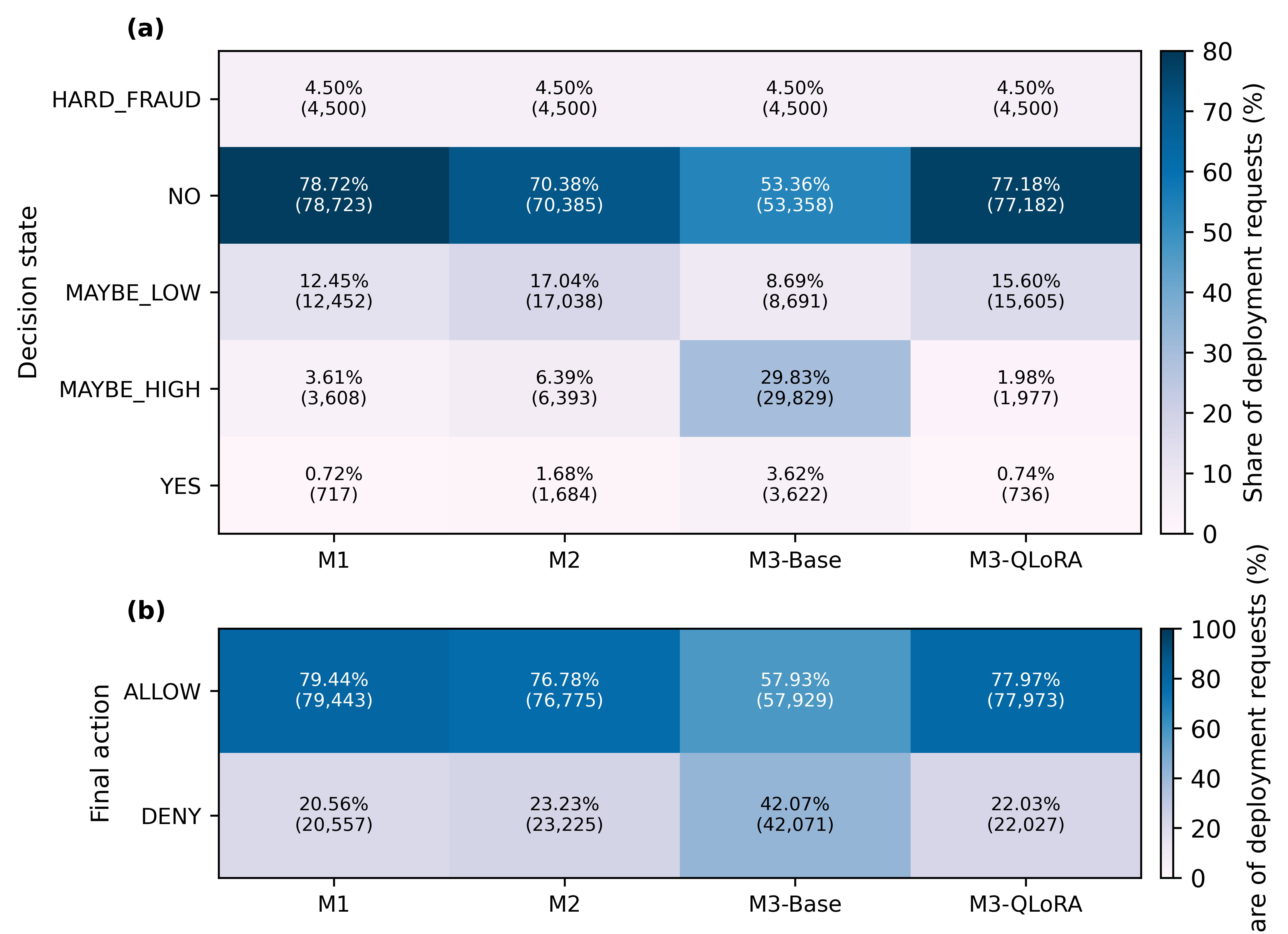}
\caption{Synthetic decision-state and final-action profiles across the four frozen upstream configurations. Panel (a) shows the percentage and count of requests assigned to each decision state, and panel (b) shows the corresponding final \plainid{ALLOW}/\plainid{DENY} outcomes. Each configuration contains 100,000 requests.}
\label{fig:synthetic-decision-profile}
\end{figure}

The controlled synthetic replay produced a complete governed decision profile for each of the four frozen upstream configurations. Figure~\ref{fig:synthetic-decision-profile} shows the decision-state distribution before final review resolution and the corresponding final \plainid{ALLOW}/\plainid{DENY} outcomes over the 100,000-request replay.

The deterministic hard gate assigned 4,500 requests (4.5\%) to hard-fraud in every configuration. The remaining differences arise from the non-hard decision streams and their frozen policy boundaries. M1 assigned 78,723 requests to \plainid{NO}, 12,452 to \plainid{MAYBE-LOW}, 3,608 to \plainid{MAYBE-HIGH}, and 717 to \plainid{YES}. M2 produced a larger review region, with 17,038 \plainid{MAYBE-LOW} and 6,393 \plainid{MAYBE-HIGH} cases. M3-Base had the largest \plainid{MAYBE-HIGH} population at 29,829 requests, while M3-QLoRA assigned 77,182 requests to \plainid{NO} and 17,582 across the two intermediate review states.

After the frozen review-resolution logic was applied, M1 produced 79,443 \plainid{ALLOW} and 20,557 \plainid{DENY} outcomes. The corresponding counts were 76,775 and 23,225 for M2, 57,929 and 42,071 for M3-Base, and 77,973 and 22,027 for M3-QLoRA. M3-Base generated a markedly larger denial workload than the other three configurations, whereas the final-action profile of M3-QLoRA was closer to those of M1 and M2.

These differences belong to the frozen upstream decision pipelines rather than to the governance mechanism. Governance-2 receives the resulting states, review requirements, and final actions and checks whether they can progress through the authorised lifecycle without changing their scientific meaning.

\subsection{Real-RF Predictive Benchmark Alignment}
\label{subsec:realrf-benchmark-results}
Each of the six study mean F1 scores fell within one published standard deviation of the corresponding mean reported in the WiFiSpectralJam matched-distribution benchmark \cite{herzalla2026wifispectraljam}. Table~\ref{tab:realrf-benchmark-alignment} compares the mean final-test F1 obtained for the six benchmark classifier families with the published all-data results. The comparison is benchmark-aligned rather than an exact reproduction because, as described in Section~\ref{subsec:realrf-models}, our protocol retains a separate development-validation partition and does not refit the models on the combined development data before final testing.

Across the six classifiers, the maximum absolute difference in mean F1 was 0.001803 and the mean absolute difference was 0.000359. Logistic Regression differed from the published mean by $-0.000158$, Random Forest by $-0.000002$, HistGradientBoosting by $-0.000057$, XGBoost by $-0.000068$, LightGBM by $+0.000066$, and the MLP by $-0.001803$. The tree and boosting families were close to the published benchmark values, while the largest discrepancy remained small and occurred for the MLP.

\begin{table}[t]
\caption{Benchmark-aligned comparison with the published WiFiSpectralJam all-data F1 results. $\Delta$F1 denotes the difference between this study and the published mean.}
\label{tab:realrf-benchmark-alignment}
\centering
\footnotesize
\renewcommand{\arraystretch}{1.08}

\begin{tabular*}{\columnwidth}{@{\extracolsep{\fill}}lccc@{}}
\toprule
\textbf{Model} &
\textbf{Published} &
\textbf{This study} &
\textbf{$\Delta$F1} \\
\midrule

LR &
$0.9721 \pm 0.0032$ &
$0.971942 \pm 0.000000$ &
$-0.000158$ \\

Random Forest &
$0.9996 \pm 0.0002$ &
$0.999598 \pm 0.000000$ &
$-0.000002$ \\

HistGB &
$0.9997 \pm 0.0002$ &
$0.999643 \pm 0.000077$ &
$-0.000057$ \\

XGB &
$0.9998 \pm 0.0003$ &
$0.999732 \pm 0.000000$ &
$-0.000068$ \\

LGBM &
$0.9998 \pm 0.0003$ &
$0.999866 \pm 0.000000$ &
$+0.000066$ \\

MLP &
$0.9948 \pm 0.0041$ &
$0.992997 \pm 0.002568$ &
$-0.001803$ \\

\bottomrule
\end{tabular*}
\end{table}

\begin{figure}[!t]
\centering
\includegraphics[width=\columnwidth]
{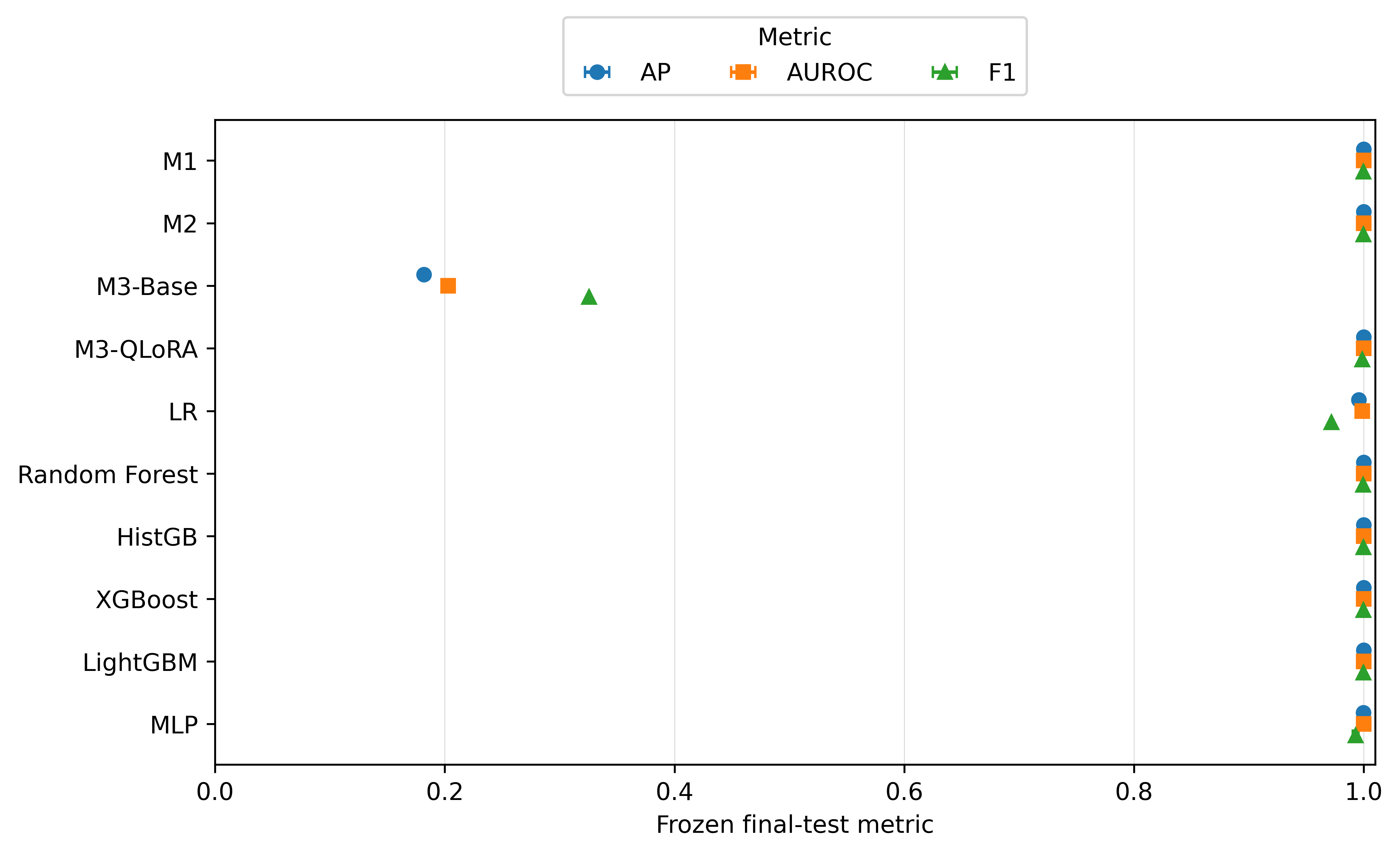}
\caption{Frozen final-test predictive performance for the real-RF upstream
systems and WiFiSpectralJam benchmark classifiers. Points show average
precision (AP), area under the receiver operating characteristic curve
(AUROC), and F1 score; error bars show the standard deviation across runs
where multiple frozen seeds are available.}
\label{fig:realrf-predictive-performance}
\end{figure}

Figure~\ref{fig:realrf-predictive-performance} places the benchmark models alongside the four frozen upstream system families using AP, AUROC, and F1 on the same sealed final-test population. M1 and M2 both achieved a mean F1 of $0.999777 \pm 0.000077$ across their three seeds. M3-QLoRA achieved an F1 of 0.998796, with AP of 0.999994 and AUROC of 0.999999. These results show that the high-performing governed streams were not dependent on a single classifier family.

M3-Base behaved differently. Its frozen final-test AP, AUROC, and F1 were 0.181750, 0.202895, and 0.325577, respectively. The review-all policy for this stream had already been fixed from development-validation evidence before the final test was opened; these final-test results did not trigger a post-hoc policy change. Governance preserved the upstream non-applicability decision and routed M3-Base cases to \plainid{REVIEW} rather than reversing the score direction or deriving new decision boundaries from the test data.

\subsection{Real-RF Governance Execution}
\label{subsec:realrf-governance-results}
The eight frozen real-RF policy streams were executed through eight independent governance-2 deployments. Each deployment processed the same 19,218 sealed final-test captures, yielding 153,744 governed stream--case executions. Of these, 134,495 cases reached a direct executable decision and were subsequently finalised, while 19,249 remained in the scientifically authorised review pending state. The aggregate workload comprised approximately 87.48\% direct cases and 12.52\% review pending cases.

The governed action totals were 108,362 \plainid{ALLOW}, 26,133 \plainid{MITIGATE}, and 19,249 \plainid{REVIEW}. Every governed case produced one initial-decision certificate, giving 153,744 decision certificates in total. Direct cases followed the four-transaction path from case submission through finalisation, whereas unresolved review cases stopped after case submission and initial-decision validation. This produced 576,478 row-level transactions across the eight deployments.

\begin{figure}[!t]
\centering
\includegraphics[width=\columnwidth]
{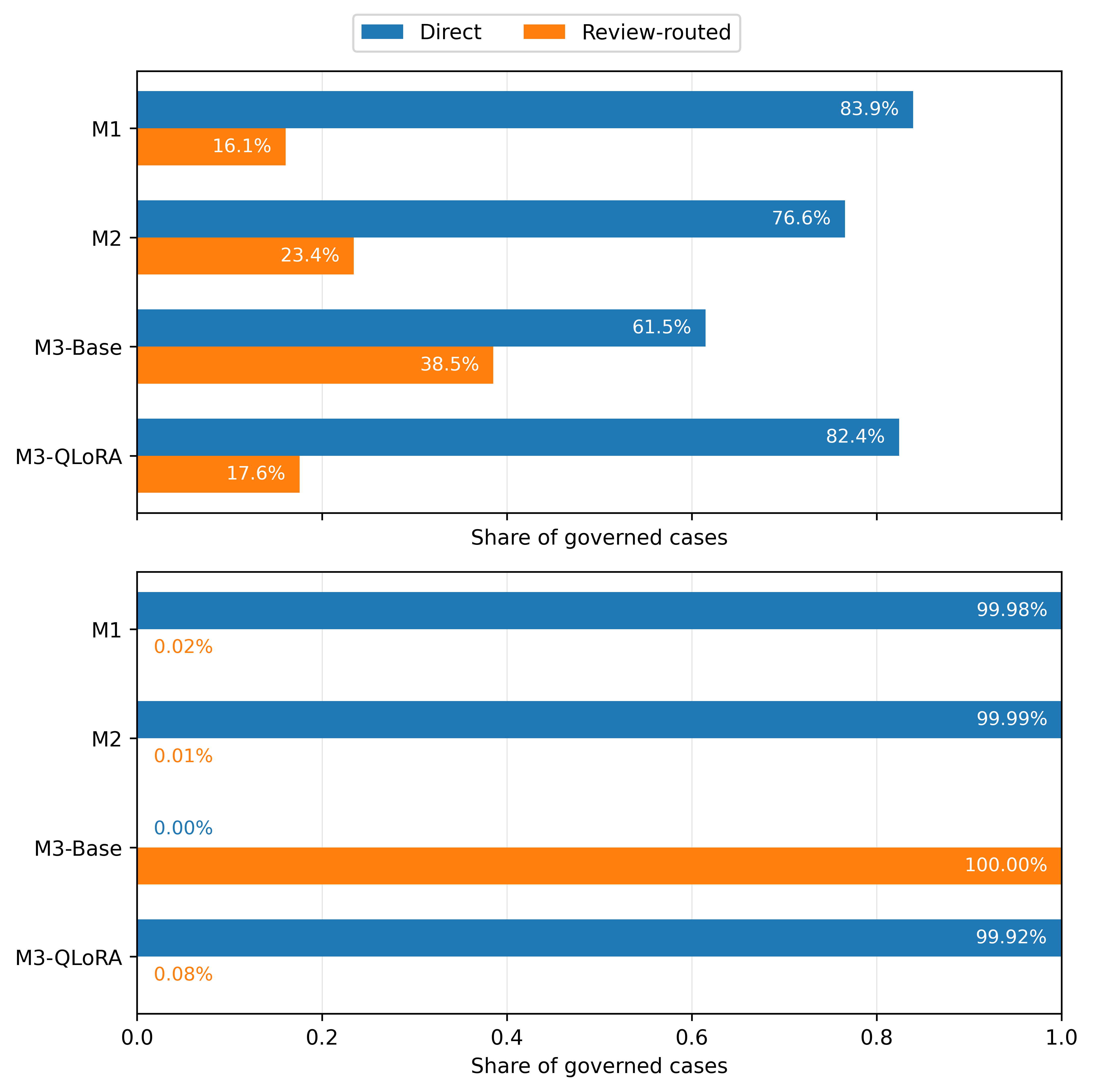}
\caption{Governance workload across the two evaluation domains. Panel (a) shows the direct and review-routed shares for the four synthetic spectrum-access configurations. Panel (b) shows the corresponding mean workload for the real-RF systems; M1 and M2 aggregate their three frozen seeds, whereas M3-Base and M3-QLoRA each represent one frozen stream.}
\label{fig:governance-workload}
\end{figure}

Figure~\ref{fig:governance-workload} compares the direct and review-routed workload produced by the synthetic and real-RF branches. In the real-RF panel, the three M1 and three M2 streams contain only a small number of review pending cases, while the M3-QLoRA stream contains 15. M3-Base is intentionally different: all 19,218 cases remain review pending because its frozen review-all policy does not authorise a direct automated decision. The figure reflects differences in upstream policy applicability rather than failures of the governance mechanism.

The complete real-RF governance run preserved the frozen scientific semantics throughout execution. Across all 153,744 cases, no state mismatch, action mismatch, certificate-integrity failure, or unexpected revert was observed. No review-resolution transaction was submitted, because no automated refiner was authorised for the real-RF branch. Review pending cases were retained without forcing a binary outcome or creating an unsupported downstream decision.

\subsection{Adversarial Enforcement Results}
\label{subsec:adversarial-results}
The full governance-2 contract rejected every deliberately invalid operation in both evaluation branches. Table~\ref{tab:adversarial-results} summarises the cross-system adversarial evaluation. Each of the four synthetic configurations rejected all 14 registered probes, with no audit failure. The independent real-RF evaluation rejected all 16 RF-adapted probes. In every case, the authoritative governance execution remained unchanged by the adversarial tests.

\begin{table}[t]
\caption{Adversarial governance enforcement across the evaluated systems.}
\label{tab:adversarial-results}
\centering
\footnotesize
\renewcommand{\arraystretch}{1.08}

\begin{tabular*}{\columnwidth}{@{\extracolsep{\fill}}lrrrr@{}}
\toprule
\textbf{System} &
\textbf{Probes} &
\textbf{Rejected} &
\textbf{Failures} &
\textbf{Unchanged} \\
\midrule
M1       & 14 & 14 & 0 & Yes \\
M2       & 14 & 14 & 0 & Yes \\
M3-Base  & 14 & 14 & 0 & Yes \\
M3-QLoRA & 14 & 14 & 0 & Yes \\
Real-RF  & 16 & 16 & 0 & Yes \\
\bottomrule
\end{tabular*}
\end{table}

For the synthetic branch, the probes exercised the eight enforcement objectives mapped in Table~\ref{tab:security-enforcement-map}. The rejected operations covered missing operational roles, substitution of otherwise active but incorrectly bound model or policy authorities, inconsistent score--state--action claims, execution of an action different from the authorised outcome, lifecycle replay, invalid automated-refiner authority, operation while paused, and use of a revoked authority. The same 14-probe registry was applied independently to each of the four synthetic governance configurations.

The real-RF audit extended the same enforcement boundary to RF-specific applicability conditions. Its 16 probes included unauthorised case and decision operations, model and policy substitution, inconsistent four-state decisions, incorrect \plainid{MITIGATE}/\plainid{ALLOW} execution, duplicate lifecycle operations, unauthorised review resolution, attempted use of an automated refiner when none was authorised, execution or finalisation of an unresolved review case, a false \plainid{NO}/\plainid{ALLOW} claim under review-all, operation while paused, and attempted use of a revoked dataset authority. All 16 invalid operations were rejected. These results establish enforcement of the tested governance boundaries; no consensus-protocol comparison was performed as part of this security evaluation.

The randomised stateful invariant test complements the fixed adversarial
registry by exercising the full contract under generated multi-step
transaction sequences. Table~\ref{tab:stateful-invariant-results} summarises
the execution scale and invariant outcomes. The absence of invariant
violations and rejected-state mutations indicates that the tested lifecycle
and governance properties remained intact across the generated sequences. 

\begin{table}[t]
\caption{Summary of randomised stateful invariant testing.}
\label{tab:stateful-invariant-results}
\centering
\footnotesize
\renewcommand{\arraystretch}{1.08}

\begin{tabular*}{\columnwidth}{@{\extracolsep{\fill}}lr@{}}
\toprule
\textbf{Measure} & \textbf{Result} \\
\midrule
Generated sequences                       & 2,000 \\
Scenario families covered                 & 10/10 \\
Sequences per scenario family             & 200 \\
State-changing transaction attempts       & 11,686 \\
Accepted transaction attempts             & 6,697 \\
Rejected invalid transaction attempts     & 4,989 \\
Invariant violations                      & 0 \\
Rejected-state mutations                  & 0 \\
\bottomrule
\end{tabular*}
\end{table}

\subsection{Capability-Ablation Results}
\label{subsec:capability-ablation-results}
The capability-ablation experiment produced a selective loss of protection for every removed enforcement family. Table~\ref{tab:capability-ablation-results} reports the security metrics for the full governance contract and the eight single-capability ablations. The full configuration rejected all 14 invalid probes, giving an IAR of 0 and SEC of 1.0. Removing any one capability reduced SEC to 0.875, corresponding to seven of the eight enforcement families remaining empirically enforced.

\begin{table}[t]
\caption{Security results for the full governance contract and single-capability ablations. ``Invalid'' reports accepted invalid probes out of 14. ``Target'' is the acceptance rate of probes assigned to the removed capability; ``Non-target'' is the rejection rate for the remaining probes.}
\label{tab:capability-ablation-results}
\centering
\footnotesize
\renewcommand{\arraystretch}{1.08}

\begin{tabular*}{\columnwidth}{@{\extracolsep{\fill}}lccccc@{}}
\toprule
\textbf{Configuration} &
\textbf{Invalid} &
\textbf{IAR} &
\textbf{SEC} &
\textbf{Target} &
\textbf{Non-target} \\
\midrule
Full           & 0/14 & 0.000 & 1.000 & --    & 1.000 \\
Role off       & 4/14 & 0.286 & 0.875 & 1.000 & 1.000 \\
Binding off    & 2/14 & 0.143 & 0.875 & 1.000 & 1.000 \\
Decision off   & 1/14 & 0.071 & 0.875 & 1.000 & 1.000 \\
Action off     & 1/14 & 0.071 & 0.875 & 1.000 & 1.000 \\
Lifecycle off  & 3/14 & 0.214 & 0.875 & 1.000 & 1.000 \\
Refiner off    & 1/14 & 0.071 & 0.875 & 1.000 & 1.000 \\
Pause off      & 1/14 & 0.071 & 0.875 & 1.000 & 1.000 \\
Revocation off & 1/14 & 0.071 & 0.875 & 1.000 & 1.000 \\
\bottomrule
\end{tabular*}
\end{table}

The ablations were target-specific. For every removed capability, the acceptance rate of its assigned probes was 1.0, while the non-target protection rate remained 1.0. Removing role enforcement admitted A01, A02, A06, and A10; removing authority binding admitted A03 and A04; removing decision consistency admitted A05; removing action enforcement admitted A07; removing lifecycle/replay protection admitted A08, A09, and A12; and the refiner-authority, pause, and authority-revocation ablations admitted A11, A13, and A14, respectively. No additional non-target probe became admissible.

The different IAR values should not be interpreted as a ranking of security importance. IAR is probe-weighted: role enforcement is represented by four probes, authority binding by two, lifecycle/replay protection by three, and each of the remaining capabilities by one. SEC avoids this weighting by assigning equal importance to the eight enforcement families. The consistent SEC value of 0.875 across the ablations reflects the loss of exactly one capability family in each experiment.

The security loss did not prevent legitimate governance execution. Each contract variant was exercised with 18 valid lifecycle controls covering all five synthetic decision states and both valid review outcomes. Every valid control was accepted for every variant, giving a VAR of 1.0 and an FRR of 0. The observed security differences arose from the deliberately removed enforcement checks rather than from a generally unusable contract variant.

\subsection{Governance Overhead}
\label{subsec:governance-overhead-results}
The local prototype measurements quantify contract-execution cost rather than production-network throughput. Table~\ref{tab:governance-overhead} reports the principal measurements for the full governance contract and the complete real-RF execution. On the 18 legitimate lifecycle controls used in the capability-ablation experiment, the Full governance contract consumed a mean of 1,023,270 gas per governed case, with a median of 1,028,894. Mean client-observed transaction-confirmation latency per complete valid case was 91.68~ms and the median was 88.41~ms.

\begin{table}[t]
\caption{Prototype governance execution measurements.}
\label{tab:governance-overhead}
\centering
\footnotesize
\renewcommand{\arraystretch}{1.08}

\begin{tabular*}{\columnwidth}{@{\extracolsep{\fill}}lr@{}}
\toprule
\textbf{Metric} & \textbf{Observed value} \\
\midrule
Full mean gas/case              & 1,023,270 \\
Full median gas/case            & 1,028,894 \\
Full mean latency/case          & 91.68 ms \\
Full median latency/case        & 88.41 ms \\
\midrule
Real-RF row transactions        & 576,478 \\
Real-RF setup transactions      & 72 \\
Real-RF total transactions      & 576,550 \\
Real-RF row gas                 & 116,322,293,625 \\
Real-RF total gas               & 116,363,967,954 \\
Sequential service-rate proxy   & 27.158 transactions/s \\
\bottomrule
\end{tabular*}
\end{table}

The large real-RF experiment provides a complementary aggregate view. Processing the 153,744 governed stream--case executions required 576,478 row-level transactions. Eight contract deployments and their role and authority setup added 72 transactions, giving 576,550 observed transactions overall. Row-level execution consumed 116,322,293,625 gas, while the inclusion of setup operations increased the total to 116,363,967,954 gas. Table~\ref{tab:governance-overhead} separates these quantities so that one-time deployment and authority-registration operations are not conflated with case-level governance processing.

Using the recorded transaction-confirmation latencies in the sequential proxy defined in \eqref{eq:service-rate-proxy} gives 27.158 transactions/s. This value is a local latency-based service-rate proxy and is not interpreted as blockchain-network throughput. The prototype uses a local Ganache execution environment and does not include a distributed validator network, consensus-throughput benchmark, or separately measured end-to-end production workload.

\section{Discussion}
\label{sec:discussion}

\subsection{Governance Across Heterogeneous Network-Security Decision Sources}
\label{subsec:discussion-heterogeneity}
The results show that heterogeneous upstream decision mechanisms can be governed through a common downstream enforcement architecture without requiring a common predictive model or identical policy semantics. The synthetic branch combines centralised, federated, and LLM-based decision sources, while the real-RF branch uses independently frozen classifiers and RF-specific decision semantics. Their scientific outputs differ substantially, as illustrated by Figures~\ref{fig:synthetic-decision-profile} and~\ref{fig:realrf-predictive-performance}, but governance-2 does not require those outputs to be generated in the same way.

Governance applies to the admissibility of a decision, not to the internal computation that produced it. Dataset, model, policy, and optional refiner authorities are bound explicitly, while score--state--action consistency and lifecycle rules are checked at the governance boundary. The architecture can consume different upstream systems while preserving their frozen semantics. Figure~\ref{fig:governance-workload} demonstrates this distinction operationally: the same governance structure accommodates both direct-decision-heavy streams and a stream such as M3-Base, for which every real-RF case remains review-routed.

The cross-domain result should not be interpreted as evidence that spectrum-access fraud decisions and RF-interference decisions are scientifically interchangeable. Their case types, negative actions, hard-gate rules, and applicability conditions remain domain-specific. The shared component is the governance mechanism: authority binding, decision validation, lifecycle enforcement, execution control, and auditable decision certification are shared while the scientific semantics remain external and frozen.

\subsection{Security Significance of Fail-Closed Enforcement}
\label{subsec:discussion-security}
The adversarial, stateful-invariant, and capability-ablation results provide
complementary evidence for the fail-closed design. The targeted probes,
summarised in Table~\ref{tab:adversarial-results}, test pre-specified
governance boundaries; the generated stateful sequences test whether lifecycle
and state invariants remain intact across multi-step interactions; and the
ablation experiment identifies which enforcement capability is responsible
when a boundary is deliberately removed.

These findings distinguish the proposed framework from prior blockchain-supported security designs by the control boundary they evaluate. Spectrum-access work has incorporated sensing into blockchain consensus \cite{fernando2023dpos}, while blockchain-based 5G resource exchange has focused on marketplace enforcement \cite{diallo2025trade5g}. Other network-security frameworks use blockchain at different stages of the security pipeline. FBBTrust uses it to persist and verify evolving trust updates in 6G networks \cite{saeedi2026fbbtrust}, whereas BLOC-SHIELD keeps a lightweight permissioned blockchain separate from federated training and inference and uses it to audit aggregation metadata in SDN intrusion detection \cite{ram2026blocshield}. Hybrid AI--blockchain systems move closer to operational control through access-decision recording \cite{alashwal2026iomt}, telecom-fraud enforcement and logging \cite{changizi2025hybrid}, and IoT IDS data protection and authority-node trust \cite{huang2026lqbids}. The present evaluation starts at a different point: an already-formed AI decision remains executable only when its dataset, model, policy, role, applicability, lifecycle, and current authority conditions are jointly satisfied. The adversarial, stateful-invariant, and ablation results provide evidence for this downstream authority-bound execution boundary, which is not evaluated through the same combination of governance properties in these prior systems.

Table~\ref{tab:capability-ablation-results} shows a highly localised effect. Every ablated contract admitted all probes assigned to the removed capability while continuing to reject every non-target probe. This pattern supports the intended decomposition of the security objectives: role checks, authority binding, decision consistency, action enforcement, lifecycle protection, refiner authority, pause control, and authority revocation each protect a distinct governance boundary. The reduction of SEC from 1.0 to 0.875 under every single-capability ablation is consistent with the removal of exactly one of the eight enforcement families.

The result is stronger than a comparison between a protected and an unprotected contract. It shows that the observed protection is not produced by a single broad rejection mechanism that happens to block many malformed calls. The individual checks contribute separable enforcement functions, and disabling one does not generally disable the others. This is particularly relevant for authority-sensitive AI governance, where a contract may receive structurally valid inputs that should still be rejected because the submitting role, model--policy binding, execution action, lifecycle state, or authority status is invalid.

\subsection{Applicability and Review Preservation}
\label{subsec:discussion-applicability}
The real-RF results also illustrate why applicability must be treated as an explicit governance condition rather than inferred from the presence of a model score. M3-Base produced a frozen score stream, but its development-validation evidence did not support a scientifically authorised four-state operating policy. The resulting review-all configuration remained review-routed throughout governance execution.

This behaviour is deliberate. A downstream governance layer could have
increased apparent decision coverage by reversing the score direction,
deriving new thresholds from the final test, or fabricating a review outcome.
Governance-2 instead preserves the upstream applicability decision, treating
unresolved review as a valid governed outcome rather than converting it into
an unsupported direct action. Figure~\ref{fig:governance-workload} shows the
resulting review workload.

For security-critical AI systems, forced automation can convert scientific uncertainty or non-applicability into an apparently complete but unsupported operational decision. In the proposed architecture, the absence of an authorised automated refinement path prevents that transition. Review is part of the governed decision semantics, not merely an intermediate state that the system is required to eliminate.

\subsection{Governance Cost and Deployment Implications}
\label{subsec:discussion-overhead}
The prototype results indicate that authority-bound enforcement introduces measurable processing cost while preserving legitimate decision execution. All legitimate control paths remained executable under the full and ablated contracts, and Table~\ref{tab:governance-overhead} reports the gas and transaction-confirmation latency associated with the full governance path. The capability-ablation results further show that, for most variants, removing an individual enforcement family changes gas consumption only modestly relative to the full contract.

For telecom and IoT networks, the relevant deployment trade-off is between the additional governance processing required before a security decision can be executed and the operational assurance gained at that boundary. Authority registration, role separation, lifecycle transitions, and decision certificates add state and transaction processing, but they allow the execution layer to reject decisions associated with invalid authorities, inconsistent policy semantics, unauthorised actors, or inadmissible lifecycle states. The practical balance will depend on the transaction volume and latency requirements of the target network service, as well as the trust and consensus model of the blockchain infrastructure used for deployment.

\subsection{Limitations and Future Directions}
\label{subsec:limitations}

The study is limited to downstream governance of already-formed AI security decisions. Governance-2 verifies registered authorities, policy bindings, score--state--action consistency, lifecycle state, and authorised execution, but it does not cryptographically attest the upstream inference process itself. The threat model also assumes correct contract execution and a trusted administrative authority; administrative-key compromise, consensus attacks, and network-level denial-of-service remain outside scope.

The blockchain experiments were conducted on a controlled local Ganache environment to isolate contract behaviour and execution cost. Reported gas, transaction-confirmation latency, and the latency-derived service-rate proxy characterise the prototype rather than a production blockchain network or consensus implementation. Evaluation on distributed permissioned or production-like infrastructures is left for future work.

The empirical branches test governance under controlled synthetic spectrum-access decisions and independently measured RF-interference data. WiFiSpectralJam provides measured RF evidence but does not constitute live telecom-fraud validation. Review-required real-RF cases are intentionally preserved in \texttt{REVIEW\_PENDING} when no authorised refiner is available; future work can evaluate authenticated human or automated review, production deployment conditions, and verifiable-inference mechanisms without changing the authority and lifecycle separation evaluated here.

\section{Conclusion}
\label{sec:conclusion}
This paper presented governance-2, an authority-bound and fail-closed governance architecture for heterogeneous AI security decisions in telecom and IoT networks. The framework separates upstream scientific decision formation from downstream operational enforcement of network security decisions and determines whether an already-formed decision is admissible for operational execution. Governed decisions are bound to registered dataset, model, policy, deployment, and optional refiner authorities, while score--state--action consistency, lifecycle validity, authority freshness, and exact-action execution are enforced without altering the scientific semantics of the upstream decision source.

Evaluation combined a controlled spectrum-access branch with an independent measured WiFiSpectralJam RF branch. Across the real-RF evaluation, 153,744 stream--case instances were processed while unresolved review states were preserved when no authorised refinement path existed. The full contract rejected every targeted invalid operation, randomised stateful testing completed 2,000 generated sequences with zero invariant violations, and capability ablation showed that removing an enforcement family exposed its assigned governance boundary while unrelated protections remained active and legitimate lifecycle paths remained executable. The results indicate that heterogeneous AI security decisions can share a common governance plane across telecom and IoT network settings while retaining domain-specific semantics and execution constraints.

\section*{Data and Code Availability}

The reproducibility materials for this study are available in the
\href{https://gitlab.surrey.ac.uk/sc02731/authority_bound_governance_of_heterogeneous-ai_security_decisions_in_telecom_iot-systems}{project GitLab repository}. The repository contains the governance-2 smart-contract implementation, synthetic and real-RF evaluation artefacts, WiFiSpectralJam processing and frozen decision streams, adversarial security tests, randomised stateful invariant tests, capability-ablation experiments, generated figures and tables, and the corresponding validation manifests.

\section*{Acknowledgment}
This work was supported by the Engineering and Physical Sciences Research Council under grant EP/Y035534/1.

\bibliographystyle{elsarticle-num}
\bibliography{references}

\end{document}